\documentclass[twocolumn,
              superscriptaddress,
              prc,
              showkeys,
              showpacs,
              nofootinbib,
              notitlepage,
              floatfix,
              ]{revtex4-2}

\usepackage[perpage]{footmisc}

\usepackage{graphicx}
\usepackage[colorlinks = true,
            linkcolor = blue,
            urlcolor  = blue, 
            citecolor = blue,
            anchorcolor = blue,
            pdftex]{hyperref}
\usepackage{xcolor}
\usepackage{amssymb}
\usepackage[nointegrals]{wasysym}
\usepackage{amsthm}
\usepackage{amsmath}
\usepackage{textcomp}
\usepackage{mathtools}
\usepackage{multirow}
\usepackage{upgreek}
\usepackage{isotope}
\usepackage{float}
\usepackage{soul, ulem}
\usepackage{lipsum} 
\usepackage{lineno}

\usepackage[separate-uncertainty,retain-explicit-plus,per-mode = symbol]{siunitx}
\usepackage{url}
\usepackage{array}
\usepackage{nicefrac}
\usepackage{comment}

\renewcommand{\arraystretch}{1.5}

\usepackage{booktabs}

\newcommand{\pmt}{PMT}

\usepackage{setspace}

\usepackage{orcidlink} 
\usepackage{currfile} 

\begin{document}

\title{XENONnT Analysis: Signal Reconstruction, Calibration and Event Selection}

\newcommand{\bologna}{\affiliation{Department of Physics and Astronomy, University of Bologna and INFN-Bologna, 40126 Bologna, Italy}}
\newcommand{\chicago}{\affiliation{Department of Physics \& Kavli Institute for Cosmological Physics, University of Chicago, Chicago, IL 60637, USA}}
\newcommand{\coimbra}{\affiliation{LIBPhys, Department of Physics, University of Coimbra, 3004-516 Coimbra, Portugal}}
\newcommand{\columbia}{\affiliation{Physics Department, Columbia University, New York, NY 10027, USA}}
\newcommand{\lngs}{\affiliation{INFN-Laboratori Nazionali del Gran Sasso and Gran Sasso Science Institute, 67100 L'Aquila, Italy}}
\newcommand{\mainz}{\affiliation{Institut f\"ur Physik \& Exzellenzcluster PRISMA$^{+}$, Johannes Gutenberg-Universit\"at Mainz, 55099 Mainz, Germany}}
\newcommand{\mpik}{\affiliation{Max-Planck-Institut f\"ur Kernphysik, 69117 Heidelberg, Germany}}
\newcommand{\munster}{\affiliation{Institut f\"ur Kernphysik, Westf\"alische Wilhelms-Universit\"at M\"unster, 48149 M\"unster, Germany}}
\newcommand{\nikhef}{\affiliation{Nikhef and the University of Amsterdam, Science Park, 1098XG Amsterdam, Netherlands}}
\newcommand{\nyuad}{\affiliation{New York University Abu Dhabi - Center for Astro, Particle and Planetary Physics, Abu Dhabi, United Arab Emirates}}
\newcommand{\purdue}{\affiliation{Department of Physics and Astronomy, Purdue University, West Lafayette, IN 47907, USA}}
\newcommand{\rice}{\affiliation{Department of Physics and Astronomy, Rice University, Houston, TX 77005, USA}}
\newcommand{\stockholm}{\affiliation{Oskar Klein Centre, Department of Physics, Stockholm University, AlbaNova, Stockholm SE-10691, Sweden}}
\newcommand{\subatech}{\affiliation{SUBATECH, IMT Atlantique, CNRS/IN2P3, Universit\'e de Nantes, Nantes 44307, France}}
\newcommand{\torino}{\affiliation{INAF-Astrophysical Observatory of Torino, Department of Physics, University  of  Torino and  INFN-Torino,  10125  Torino,  Italy}}
\newcommand{\ucsd}{\affiliation{Department of Physics, University of California San Diego, La Jolla, CA 92093, USA}}
\newcommand{\wis}{\affiliation{Department of Particle Physics and Astrophysics, Weizmann Institute of Science, Rehovot 7610001, Israel}}
\newcommand{\zurich}{\affiliation{Physik-Institut, University of Z\"urich, 8057  Z\"urich, Switzerland}}
\newcommand{\paris}{\affiliation{LPNHE, Sorbonne Universit\'{e}, CNRS/IN2P3, 75005 Paris, France}}
\newcommand{\freiburg}{\affiliation{Physikalisches Institut, Universit\"at Freiburg, 79104 Freiburg, Germany}}
\newcommand{\napels}{\affiliation{Department of Physics ``Ettore Pancini'', University of Napoli and INFN-Napoli, 80126 Napoli, Italy}}
\newcommand{\nagoya}{\affiliation{Kobayashi-Maskawa Institute for the Origin of Particles and the Universe, and Institute for Space-Earth Environmental Research, Nagoya University, Furo-cho, Chikusa-ku, Nagoya, Aichi 464-8602, Japan}}
\newcommand{\laquila}{\affiliation{Department of Physics and Chemistry, University of L'Aquila, 67100 L'Aquila, Italy}}
\newcommand{\tokyo}{\affiliation{Kamioka Observatory, Institute for Cosmic Ray Research, and Kavli Institute for the Physics and Mathematics of the Universe (WPI), University of Tokyo, Higashi-Mozumi, Kamioka, Hida, Gifu 506-1205, Japan}}
\newcommand{\kobe}{\affiliation{Department of Physics, Kobe University, Kobe, Hyogo 657-8501, Japan}}
\newcommand{\kit}{\affiliation{Institute for Astroparticle Physics, Karlsruhe Institute of Technology, 76021 Karlsruhe, Germany}}
\newcommand{\tsinghua}{\affiliation{Department of Physics \& Center for High Energy Physics, Tsinghua University, Beijing 100084, P.R. China}}
\newcommand{\ferrara}{\affiliation{INFN-Ferrara and Dip. di Fisica e Scienze della Terra, Universit\`a di Ferrara, 44122 Ferrara, Italy}}
\newcommand{\groningen}{\affiliation{Nikhef and the University of Groningen, Van Swinderen Institute, 9747AG Groningen, Netherlands}}
\newcommand{\westlake}{\affiliation{Department of Physics, School of Science, Westlake University, Hangzhou 310030, P.R. China}}
\newcommand{\shenzhen}{\affiliation{School of Science and Engineering, The Chinese University of Hong Kong, Shenzhen, Guangdong, 518172, P.R. China}}
\newcommand{\coimbrapoli}{\affiliation{Coimbra Polytechnic - ISEC, 3030-199 Coimbra, Portugal}}
\newcommand{\uniheidelberg}{\affiliation{Physikalisches Institut, Universit\"at Heidelberg, Heidelberg, Germany}}
\newcommand{\roma}{\affiliation{INFN-Roma Tre, 00146 Roma, Italy}}
\newcommand{\bucknell}{\affiliation{Department of Physics \& Astronomy, Bucknell University, Lewisburg, PA, USA}}
\author{E.~Aprile\,\orcidlink{0000-0001-6595-7098}}\columbia
\author{J.~Aalbers\,\orcidlink{0000-0003-0030-0030}}\groningen
\author{K.~Abe\,\orcidlink{0009-0000-9620-788X}}\tokyo
\author{S.~Ahmed Maouloud\,\orcidlink{0000-0002-0844-4576}}\paris
\author{L.~Althueser\,\orcidlink{0000-0002-5468-4298}}\munster
\author{B.~Andrieu\,\orcidlink{0009-0002-6485-4163}}\paris
\author{E.~Angelino\,\orcidlink{0000-0002-6695-4355}}\torino\lngs
\author{J.~R.~Angevaare\,\orcidlink{0000-0003-3392-8123}}\nikhef
\author{D.~Ant\'on~Martin\,\orcidlink{0000-0001-7725-5552}}\chicago
\author{F.~Arneodo\,\orcidlink{0000-0002-1061-0510}}\nyuad
% ...
\collaboration{XENON Collaboration}\email[]{xenon@lngs.infn.it}\noaffiliation

\date{\today}

\begin{abstract}
\noindent 
The XENONnT experiment, located at the INFN Laboratori Nazionali del Gran Sasso, Italy, features a 5.9\;tonne liquid xenon time projection chamber surrounded by an instrumented neutron veto, all of which is housed within a muon veto water tank. Due to extensive shielding and advanced purification to mitigate natural radioactivity, an exceptionally low background level of (15.8 $\pm$ 1.3)\;events/(tonne$\cdot$year$\cdot$keV) in the (1, 30)\;keV region is reached in the inner part of the TPC. XENONnT is thus sensitive to a wide range of rare phenomena related to Dark Matter and Neutrino interactions, both within and beyond the Standard Model of particle physics, with a focus on the direct detection of Dark Matter in the form of weakly interacting massive particles (WIMPs).
From May 2021 to December 2021, XENONnT accumulated data in rare-event search mode with a total exposure of one tonne $\cdot$ year. This paper provides a detailed description of the signal reconstruction methods, event selection procedure, and detector response calibration, as well as an overview of the detector performance in this time frame. 
This work establishes the foundational framework for the `blind analysis' methodology we are using when reporting XENONnT physics results.
\end{abstract}

\graphicspath{{section_2/plots/}{section_3/plots/}{section_4/plots/}}

\maketitle

\section{Introduction}
\label{sec:intro}

A wide range of astrophysical and cosmological observations indicate a universe where only a small portion of the matter is baryonic and the total mass content is dominated by a new, yet unknown, form of non-luminous substance, or dark matter~\cite{Bertone:2004pz}. Several candidate particles, not contained in the Standard Model of particle physics, can solve the dark matter problem, and one of the most intriguing are the weakly interacting massive particles (WIMPs)~\cite{Roszkowski:2017nbc}. In recent years, axions~\cite{Marsh:2015xka} and bosonic dark matter~\cite{PhysRevD.78.115012}, such as dark photons, have gathered a lot of attention in the scientific community. Nevertheless, WIMPs are still one of the main search candidates in the quest to identify dark matter. In the context of direct detection of dark matter, dual-phase xenon time projection chambers (TPCs) are at the forefront of probing WIMPs in the GeV to TeV mass range~\cite{xenon_nt_wimp.131.041003, PandaX-4T:2021bab, LZ:2022lsv}, with the XENON project being a key contributor to this effort.

Together with Ref.~\cite{xenonnt_analysis_2}, this paper reports on the analysis methods employed for the main physics analyses of the first science run of XENONnT (SR0)~\cite{xenonnt_lower, xenon_nt_wimp.131.041003}, focusing on the techniques of signal reconstruction, event selection, and detector calibration.

The XENONnT experiment is the present stage of the XENON project. Most of the service systems and infrastructure have been inherited from its predecessor, the XENON1T experiment~\cite{aprile2017xenon1t}. The XENONnT experiment is located in Hall B of the INFN Laboratori Nazionali del Gran Sasso (LNGS), Italy, consisting of three nested detectors. A water tank with 10\;m diameter and height functions as a Cherenkov muon veto (MV). Next to the water tank, a 3-floor service building hosts the experiment’s infrastructure. Within the water tank, a neutron veto (NV)~\cite{xenonnt_nv} is installed, which consists of an inner region enclosed by reflective panels, equipped with 120 Hamamatsu R5912-100-10 8” PMTs. The TPC is placed at the center of the water tank inside a double-walled stainless-steel cryostat. Twenty-four Polytetrafluoroethylene (PTFE) reflector panels surround the TPC volume, marking the boundary of the TPC active volume with a 1.34\;m diameter which is filled with 5.9\;tonnes of liquid xenon (LXe), in a double-walled vacuum-insulated cryostat. Two arrays of 3” Hamamatsu R11410-21 photomultiplier tubes (PMTs)~\cite{PMT_XENONnT:Antochi2021}, totaling 494 units, are placed at the top and the bottom of the cylinder. 

An electric drift field is generated in the LXe by a cathode placed 60\;mm above the bottom PMT array and a gate electrode. These two electrodes, separated by 1486\;mm at LXe temperature, demarcate the active region. The detector is filled with LXe to 5.0\;mm above the gate electrode, above which xenon exists in gaseous form. An anode electrode is placed 8\;mm above the gate to establish an extraction field across the liquid-gas interface. Two screening electrodes at 5.3\;mm above the bottom and 40.7\;mm below the top PMTs protect the photosensors from the electric fields. Two and four additional wires, for the gate and the anode, respectively, are installed perpendicular to the other wires to minimize the effect of gravitational and electrostatic sagging. These wires are referenced as the perpendicular wires, and create localized variations in signal properties. Two concentric sets of oxygen-free copper field-shaping rings ensure the uniformity of the drift field and minimize charge-insensitive regions. Both inner and outer cryostats have domed upper sections with several access ports connecting the TPC to the rest of the infrastructure. Additional information can be found in Ref.~\cite{xenonnt_instrumentation, XENON:xenonnt_field_cage}.

Like other noble liquids, xenon responds to energy depositions in the form of atom recoils, resulting in atom ionizations and excitations. Excited xenon atoms combine to form excited dimers whose dissociations emit 175\;nm scintillation photons~\cite{175nm:FUJII2015293} which are measured by the PMTs and are referred to as S1 signal. 
The free electrons produced in the ionization channel are displaced from the interaction site by the electric drift field toward the liquid-gas interface at the top of the TPC. At this interface, the much stronger extraction field extracts the electrons into the gas phase and creates the proportional scintillation S2 signal~\cite{LANSIART197647}. The partial recombination of free electrons with ions also forms excited dimers, whose dissociation contributes to the S1 signal. The splitting of energy between ionizations and excitations results in an anti-correlation between the S1 and S2 signals. The S2/S1 signal ratio enables the differentiation between electronic recoils (ER) and nuclear recoils (NR). NR events, which include expected signals from WIMPs~\cite{xenon_nt_wimp.131.041003}, and coherent elastic neutrino-nucleus scattering~\cite{xenonnt_cevns_2024}, and neutrons as a source of background, are of primary interest. Conversely, ER events, caused by $\beta$ particles, gamma radiation and neutrino-electron scattering, constitute the predominant background in the search for WIMPs. However, ER events can also be used to probe new physics, such as axions~\cite{xenonnt_lower}, and double weak decay searches~\cite{PhysRevC.106.024328}.

The XENONnT experiment was installed in 2020 and commissioned by Spring 2021. The first scientific data acquisition periods, referred to as SR0, is detailed in Sec.~\ref{sec:2}. Sec.~\ref{sec:3} describes the data processor used to convert the raw data obtained during this period into physical quantities and properties of S1 and S2 signals, and the event simulation framework used to evaluate the performance of the data processor. The processor also reconstructs the interaction position of each event (Sec.~\ref{sec:4}) and applies corrections to the measured signals to account for spatial and temporal dependencies (Sec.~\ref{sec:5}). The leading ER and NR searches are based on selecting a clean sample of single-scatter events (Sec.~\ref{sec:6}) inside a central fiducial volume with a reduced background level. Finally, the energy reconstruction performance is presented in Sec.~\ref{sec:7}.

\section{Detector operation and stability}
\label{sec:2}

\subsection{First Science Run}
\label{subsec1:Sciences_runs}

\begin{figure}[t!]
  \centering
  \includegraphics[width=\linewidth]{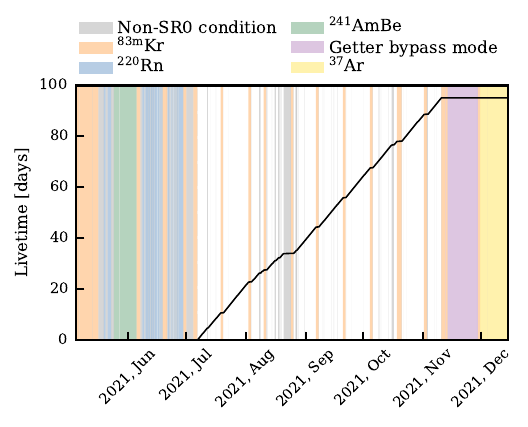}
  \caption{Live time development of XENONnT SR0. The solid black line shows the cumulative science data without deadtime correction. Colored bands highlight calibration periods and intervals of detector conditions unsuitable for scientific analysis from PMT trips, hotspot periods, and maintenance operations.}
  \label{fig:dect_ope:science_run}
\end{figure}

The datasets recorded during SR0 between July 6$^{\text{th}}$ and November 10$^{\text{th}}$ 2021 include physics-data, with a total live time of 97.1 days and used for rare-event-search analyses~\cite{xenonnt_lower, xenon_nt_wimp.131.041003}, as well as calibration data performed before, during, and after this period~\cite{xenonnt_instrumentation}. The SR0 data-taking campaign is shown in Fig.~\ref{fig:dect_ope:science_run}. The following types of calibration sources are used to quantify the detector response to ionising radiation: \isotope[83\text{m}]{Kr}, \isotope[241]{AmBe}, \isotope[220]{Rn} and \isotope[37]{Ar}. 

\isotope[83\text{m}]{Kr} atoms were injected through the gas xenon recirculation path into the LXe TPC volume about every two weeks. \isotope[83\text{m}]{Kr} decays via subsequent emission of 32.1\;keV and 9.4\;keV conversion electrons, with half-lives of 1.83\;h and 157\;ns respectively~\cite{NNDC}. The first decay is slow enough that the source distributes uniformly in the detector after injection, as shown by the distribution of the reconstructed positions of \isotope[83\text{m}]{Kr}. The second decay provides a signature of two subsequent S1 signals, reconstructed as either a merged S1 peak (41.5\;keV) or two separate S1 peaks (32.1\;keV and 9.4\;keV). Most of the S2 signals are not separable because S2s have $\mathcal{O}(1)$\;$\upmu$s widths. $^{83\text{m}}$Kr events were used to monitor and characterize spatial and temporal variations of detector response at those energies.

An \isotope[241]{AmBe} source, inserted in the water tank and deployed in different positions around the outer cryostat, was used to characterize the TPC response to NR events and evaluate the neutron veto detection efficiency.

Two calibration sources, \isotope[220]{Rn} and \isotope[37]{Ar}, were used to characterize the ER response. The first provided a continuous ER spectrum at low energies thanks to its $\upbeta$ emitter daughter \isotope[212]{Pb}, and it was used for ER band modeling and to develop data selection criteria. The decay of \isotope[37]{Ar}, with a 35 days half-life, leads to low-energy events at 2.82\;keV and 0.27\;keV via (K- and L- shell) electron capture~\cite{XENON1T:Ar37-calibration}. These were primarily used to further understand the detector response near the energy threshold. To avoid unwanted \isotope[37]{Ar} contamination in the physics search data, the source was injected at the end of the SR0 and removed afterward via cryogenic distillation~\cite{XENON:2021fkt}.

Besides the physics search and detector calibrations, the detector was operated in different conditions, e.g., PMTs gain calibration and anode-ramped down periods due to strong single electron emission. These periods are excluded from SR0 and marked gray in Fig.~\ref{fig:dect_ope:science_run}. 

To investigate the XENON1T low-energy ER excess~\cite{lowER:Aprile:2020tmw}, a test was conducted by modifying the xenon recirculation scheme to potentially enhance tritium sources. This adjustment involved bypassing a GXe getter upstream of the Radon Removal System~\cite{xenonnt_instrumentation}, thereby potentially increasing the concentration of water, tritiated water, and tritiated hydrogen within the xenon target. It is speculated that tritium could account for the observed ER excess. As a result, this specific data-acquisition mode, referred to as getter bypass mode, aimed to understand the potential impact of tritium contamination. However, the test results indicated no significant increase in tritium level.

\subsection{Detector Conditions}
\label{subsec1:detect_conditions}

During the commissioning phase of the detector, the field strengths within the different TPC regions were optimized to achieve the best possible detection efficiency and ER/NR discrimination. However, following a cathode short-circuit event, the cathode and bottom screening electrodes were shorted. As a result, the drift field in the active volume had to be reduced to 22.9~$_{-0.3}^{+0.5}$\;V/cm (uncertainties reflect the standard deviation across the volume), resulting in a long maximum electron drift time of 2.2 ms. Furthermore, sporadic and localized high rates of single electron events (hot spots) limited the extraction field intensity to 2.9\;kV/cm. The mitigation of these hot spots required a few temporary shutdowns of the top electrode stack and, consequently, an interruption of the data acquisition. The voltage steps between successive field shaping electrodes were optimized by setting the independent power supply of the top field shaping wire electrode to +0.65\;kV to reduce field inhomogeneity and the charge-insensitive mass inside the TPC volume~\cite{XENON:xenonnt_field_cage}.

During the entire science run, the detector was operated under stable thermodynamic conditions with average detector pressure, liquid xenon temperature, and liquid-gas interface level of $(1.890\pm0.004)$\;bar, $(-97.15\pm0.4)$\;°C and $(5.0\pm0.2)$\;mm, respectively.

The cryogenic distillation campaigns, conducted during the detector commissioning for krypton and continuously during SR0 for radon, reduced the $^{\text{nat}}$Kr/Xe molar concentration to $(56\pm36)$ ppq and the $^{222}$Rn level to $(1.87\pm0.09)$ $\upmu\text{Bq/kg}$~\cite{xenonnt_radon_2024}. Similarly, evacuating the detector for a period of 3 months following the cryostats' sealing helped to reduce the water content by minimizing outgassing. The water concentration of the vaporized liquid xenon circulating from the cryostat measured during SR0 was consistently below the sensor sensitivity of 0.5\;ppb (mol/mol). Finally, the upgraded gas and new liquid xenon purification lines lowered the concentration of electronegative impurities, e.g., O$_2$, to a level such that electron lifetimes\footnote{Defined as the mean survival time for a free electron in the detector before it is attached to an impurity. Sec.~\ref{subsec4:s2_spatial_cor} for additional information.} consistently exceed $\sim$10\;ms during SR0.

\begin{figure}[h]
  \centering
  \includegraphics[width=\linewidth]{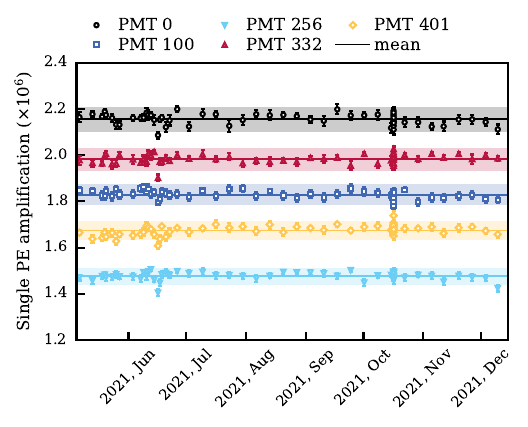}
  \caption{Evolution of single photoelectron amplification during SR0. The dense segment in June corresponds to radon calibration, involving multiple PMT calibration sessions. The high-density region around October corresponds to a dedicated calibration campaign focusing on evaluating the systematic uncertainty in gain computation. The displayed PMTs reflect the observed behavior in SR0, demonstrating, on average, stable gains with fluctuations confined to 2.5\% highlighted by colored bands.}
  \label{fig:dect_ope:PMT_gain_stability}
\end{figure}

The voltages supplied to the photosensors were individually optimized during the commissioning phase to minimize the afterpulses rate as well as spurious light emissions while keeping a uniform single photoelectron (PE) acceptance at the digitizer threshold (typically 15 ADC counts) equal to $(91.2 \pm 0.2)$\%. This configuration was achieved with an average PMT gain equal to $(1.87\pm0.35)\times10^6$, where the reported uncertainty reflects the standard deviation over all PMTs, and a maximum bias voltage limited to -1.5\;kV. The PMT gains were determined at least once a week by flashing LEDs~\cite{xenonnt_instrumentation} and using the analysis method from XENON1T~\cite{Aprile:2019bbb}. 
Fig.~\ref{fig:dect_ope:PMT_gain_stability} illustrates the gain trends during SR0 of five stable PMTs (numbered 0, 100, 256, 332, 401), which are indicative of the behavior observed in the majority of PMTs within the XENONnT TPC. Modeling the gain evolution is based on a linear fit of successive sub-sets of adjacent gain calibration data points after smoothing the latter with a custom Savitzky-Golay filter~\cite{1964AnaCh..36.1627S}. The deviation between the measured and modeled gains consistently remained below $\sim 2.5$\% for approximately 93\% of all sensors.

A total of 17 out of 494 PMTs were excluded from the SR0 analysis due to high electronic noise (1), unstable behavior (1), light emission (2), increasing afterpulse rate (11), and damage to the cable connection (2).

The stability of S1 and S2 signals over time is ensured by regularly monitoring the evolution of light and charge yields (LY and CY), discussed in Sec.~\ref{sec:7}, using data from mono-energetic sources spanning from 9.4\;keV (\isotope[83\text{m}]{Kr}) to 5.6\;MeV ($\upalpha$ from \isotope[222]{Rn}). Throughout blinded data acquisition, the LY and CY values demonstrated remarkable stability, with their deviations from the mean not exceeding 1\% and 1.9\%, respectively, as detailed in~\cite{pierr2022}.

\section{Signal reconstruction and simulation}
\label{sec:3}

\begin{figure*}[ht!]
    \centering
    \includegraphics[width=\textwidth, scale=0.5]{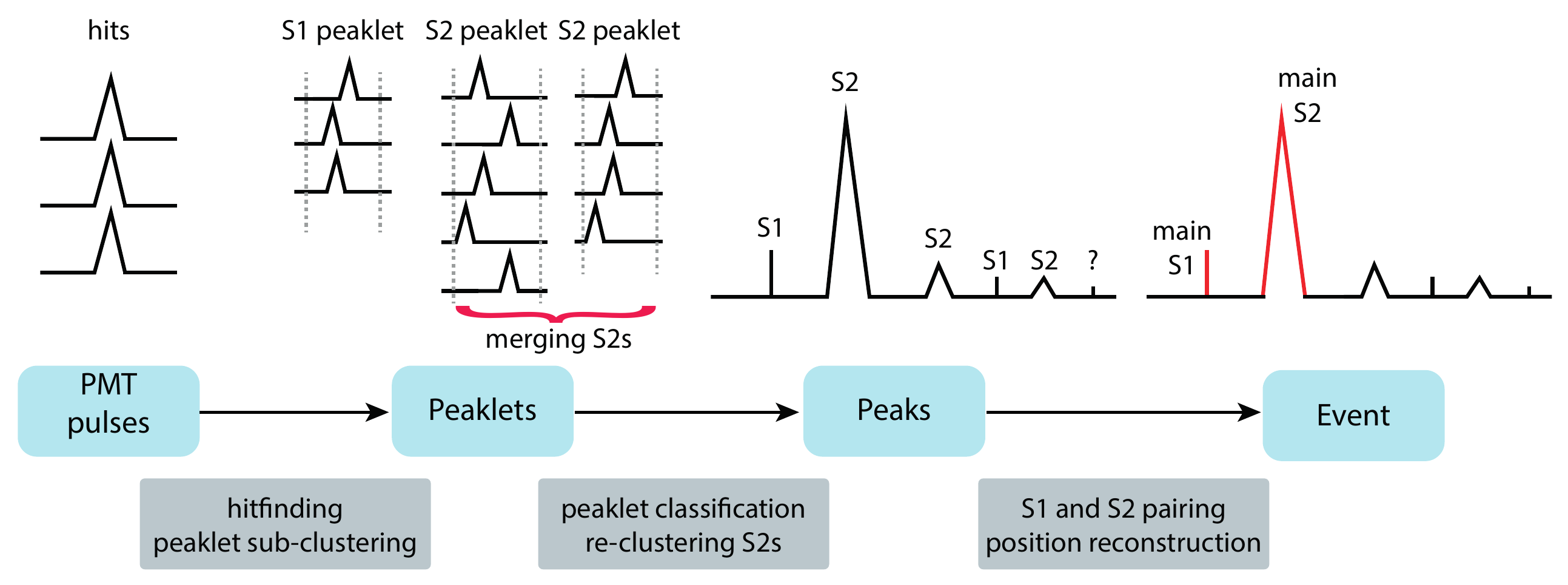}
    \caption{
        The working principle applied in \texttt{straxen} to build (S1, S2) events starting from PMT pulses and passing through intermediate objects (blue tiles). Reconstructed peaklets which do not satisfy either S1 signal classification or S2 signal classification are represented in the schema by the ``?'' label. The main processing steps are reported (grey tiles). Diagram adapted from~\cite{giovanni_volta_thesis_2023} and  from~\cite{Wittweg}.
        }
    \label{fig:reconstruction_chain}
\end{figure*}

The scintillation light of the S1 and S2 signals liberates photoelectrons from the PMT photocathodes, which create a pulse that is digitized by the triggerless XENONnT DAQ~\citep{XENON:2022vye}. Each signal that passes a channel-dependent threshold is digitized at a sampling rate of 100\;MHz. The entire stream of data from the PMTs is stored on disk long-term without further trigger, processing, or triage of the raw data. This enables us to reprocess the data with new algorithms or improved detector understanding at any given time. The reconstruction chain, shown in Fig.~\ref{fig:reconstruction_chain}, aims to extract and match the S1 and S2 signals from the data stream.

\subsection{Reconstruction Chain}
\label{subsec:rec_chain}

The reconstruction algorithms search for signals in the \pmt{} waveforms. The time intervals of these signals are called ``hits'' and are defined as the time interval above threshold extended by a window of 30\;ns on the left and 200\;ns on the right.
These per-\pmt{} hits are sequentially grouped with neighboring hits (from any \pmt{}) into clusters, where the time gap between consecutive hits within a cluster is \SI{700}{\ns} or less.
Isolated hits (``lone hits''), which have no neighboring hits in this time window, are primarily due to afterpulses or dark counts and are handled and stored separately.
The clustered groups of hits are iteratively split into sub-clusters based on their timing information and the summed waveform of all hits in the cluster using a natural break algorithm~\cite{strax}. This splitting is necessary to separate S1 signals from PMT afterpulses or nearby peaks. Sub-clusters exhibiting saturation are corrected based on the method developed in XENON1T~\cite{Aprile:2020yad} using a pulse model built from the non-saturated channels.

\begin{figure}[h!]
    \centering
    \includegraphics[width=\linewidth]{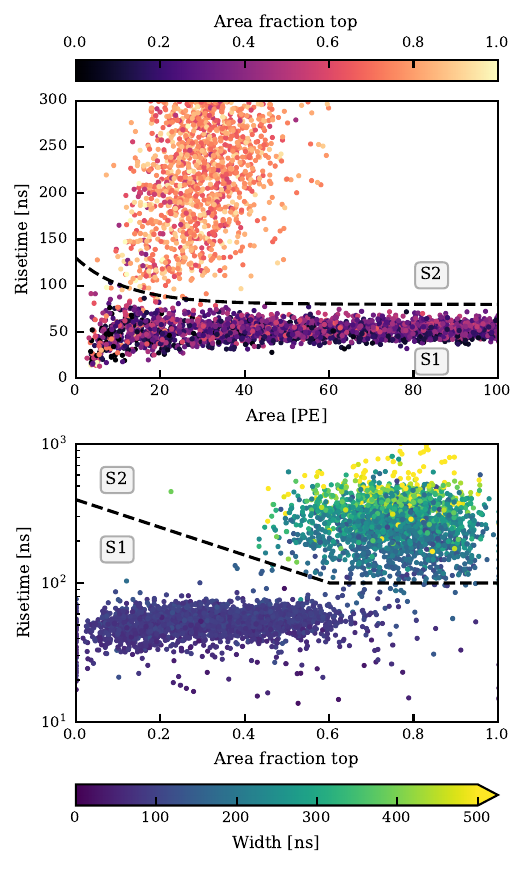}
    \caption{Classification boundaries in \texttt{straxen}~\citep{straxen} between S1 and S2 signals using the peaklets characteristics, showing the S1/S2 classification boundaries in area versus rise time (top) and area fraction top versus rise time (bottom). The data shown are S1 peaks from a \isotope[220]{Rn} calibration and isolated single electrons (the smallest S2 signals). S1 signals are typically much faster (i.e., have a shorter rise time) than secondary scintillation; this property is used in the top panel to discriminate between the two signals. S1 signals are expected to be measured predominantly by the bottom \pmt{} array, while secondary scintillation light is produced in the xenon gas and, therefore, detected mostly by the top array. Furthermore, to minimize the number of fake S1 signals from accidental coincidences between PMT dark counts, a minimum of three PMTs must contribute to the tight coincidence window for the peaklet to be classified as S1 signal. Figure adapted from~\cite{Angovaare2023}.}
    \label{fig:classification_boundaries}
\end{figure}

These sub-clusters, called ``peaklets'', are sequentially classified as S1 or S2 peaklets based on their waveform shapes along the classification boundaries shown in Fig.~\ref{fig:classification_boundaries}. These boundaries are encoded in the reconstruction software \texttt{straxen}~\citep{straxen}. The boundaries and classification utilize several characteristics of the peaklets:
\begin{itemize}
    \item The \textit{area} of the peaklet, which is the total charge (gain corrected, in PE) measured by all \pmt{}s during the peaklet.
    \item The \textit{rise time} is defined as the time between the 10\% and the 50\% area quantiles of the sum waveform.
    The 10\% and 50\% area quantiles are obtained from the time intervals commencing at the start of the first contributing hit, wherein 10\% and 50\% of the total charge of the peaklet is achieved, respectively.
    \item The \textit{width} of the peaklet is the time range where the central 50\% of the area of the peak resides.
    \item The \textit{tight coincidence} (TC) is the number of different \pmt{}s that have a hit within $\pm$\SI{50}{\ns} around the time of the peaklet's maximal amplitude.
    \item The \textit{area fraction top} (AFT) is the fraction of the total area seen in the top \pmt{} array.
\end{itemize}

S2 signals originating from the bottom of the detector exhibit a larger temporal spread due to longitudinal diffusion during the electron cloud drift. Thus, S2 peaklets from a few electron signals may have been mistakenly split during the first stage of the peaklet building. Consequently, a merging step is applied to S2 peaklets using a gap-size clustering~\citep{straxen}. Adjacent S2 peaklets are merged until the combined duration exceeds \SI{50}{\us} or no further candidates are nearby. This cap prevents the inclusion of secondary delayed electrons and photo-ionization electrons, which are additional electrons freed via photoemission, into the main S2 peak. The allowable gap size for merging depends on the integrated peak area of the resulting S2. If any lone hits fall within the duration of the newly merged S2 peaklet, they are also included to avoid depth-dependent area bias for small S2 signals. Following this step, all merged S2 peaklets, along with unmerged peaklets such as S1s or S2s without a merging partner, are referred to as ``peaks''.

After defining S1 and S2 peaks from the PMT waveforms, the reconstruction algorithms build events from peaks. As there is no global trigger enforced at the DAQ~\citep{XENON:2022vye}, there is no predetermined event definition (in contrast to XENON1T~\citep{XENON:2019bth}). Events are built from the stream of peaks and are defined as the time region spanned around S2 signals with an area $\leq 100$\;PE, called the ``triggering peak''. Additionally, the triggering peak must have fewer than eight other neighboring peaks in a window of \SI{\pm10}{\ms} that have $>50\%$ of the area of the triggering peak (called the ``n\_competing'' requirement). This requirement ensures that only the largest S2 signals act as triggering peaks and that, e.g., continued photo-ionization tails and delayed electrons after large S2 signals do not lead to a high number of triggering peaks. Events are defined as the time window encompassing \SI{2.45}{\ms} prior to and \SI{0.25}{\ms} after the triggering peak. If multiple event windows overlap, they are merged. The primary S1 is identified by the largest S1 peak within the event window, whereas the main S2 peak is determined by the largest S2 signal detected after the main S1. The second largest S1(S2) peak within an event are designated as ``alternative" S1(S2). Similarly, the alternative S2 must be recorded following the main S1. Identifying these alternative peaks is crucial for recognizing multi-scatter events.

The n\_competing requirement leads to an energy-dependent event-building efficiency. For the S2-area threshold of 500\;PE used for the ER search~\cite{xenonnt_lower}, the event-building efficiency is 99.3\%. For the WIMP search~\cite{xenon_nt_wimp.131.041003}, the threshold is 200\;PE, corresponding to an efficiency of 97.2\%. These efficiencies are accounted for in the inference~\citep{xenonnt_analysis_2}.

\subsection{Signal Simulation}

Monte Carlo (MC) simulations of the detector response to S1 and S2 signals are performed to understand potential biases and inefficiencies. The simulation workflow, starting from the photon and electron yields up to DAQ simulation, is managed using the waveform simulator (\texttt{WFSim}) package~\citep{peter_gaemers_2022_7216324}.\\

The simulation workflow can integrate with the XENONnT simulation package~\cite{xenonnt_mc:Aprile_2020} based on the \texttt{Geant4} toolkit~\cite{ALLISON2016186, AGOSTINELLI2003250} to generate the necessary energy depositions. The modeling of liquid xenon response to the deposited energy is the first step of the simulation pipeline. The \texttt{epix} (electrons and photons instructions for XENON) package~\cite{epix} processes energy depositions to evaluate primary scintillation photons and electrons yield per interaction, using models from the Noble Element Simulation Technique (\texttt{NEST}) software~\cite{Szydagis:2020isq}. 

The light quanta derived from \texttt{epix} are used to generate detected photon arrival times per PMT necessary to simulate S1 signals. The photon scintillation delay times are generated using \texttt{NEST}'s. The per-PMT hit distribution as well as the time delay from optical propagation are then produced based on a pre-computed probability maps obtained from optical simulation. These optical simulations are performed with \texttt{Geant4} using a TPC model whose optical properties are derived by matching the MC simulation to $^{83\text{m}}$Kr calibration data, as detailed in~\citep{giovanni_volta_thesis_2023}.

To simulate S2 signals, the first step consists of modeling the electron drift using pre-computed electric field map~\cite{XENON:xenonnt_field_cage}. The processes of electron diffusion and losses to electronegative impurities are also accounted for when evaluating the number, time, and position of electrons reaching the gate electrode. The probability of an electron being extracted from liquid xenon to the gas amplification phase is derived from a data-driven map. For each electron extracted in the gas phase, we estimate the number of photons detected, depending on its position, arrival time in the gas, and the measured secondary gain (PE measured per electron drifting in the gas phase). As for S1 simulations, the detected photon arrival times per PMT are obtained from optical simulations. The final photon hit times are generated by summing the initial electron arrival time with a sampled atomic excitation time, scintillation delay, and optical propagation delay.

For each of the generated PMT hits, the PMT and DAQ readout responses are applied, accounting for pulse shape and amplification with data-driven templates, digitizer threshold value, sampling rate, and electronic noise from pre-recorded samples. This results in simulated pseudo-data in the same format as provided by our real detector so that the same data processing workflow can be used for its analysis. More details about this waveform emulating framework can be found in~\citep{Ramirez_diego_thesis_2022}. This full-chain simulation workflow is used to evaluate reconstruction efficiency and its bias, to complement the information extracted from calibration data. 

\subsection{Reconstruction Efficiency}
\label{subsec:rec_eff}

The reconstruction efficiency is the probability that a given peak is reconstructed with the proper classification. The main mechanism of S1 reconstruction efficiency loss arises from the S1 classification (Fig.~\ref{fig:classification_boundaries}). For the first XENONnT analysis in SR0, as stated in Sec.~\ref{subsec:rec_chain}, hits in at least three distinct PMTs within the TC window of the peak are required. This criterion is particularly effective at low energies, where it significantly reduces accidental coincidence events caused by dark counts or electronic noise misidentified as S1 signal and paired with a lone S2 signal. The second constraint impacting the accurate classification of S1 signals involves the rise time and AFT boundaries, designed to distinguish between S1 signals and single electrons (SE), as shown in Fig.~\ref{fig:classification_boundaries}. Since S1 photons often undergo total reflection at the liquid-gas interface, S1 signals originating from the upper part of the LXe volume exhibit a broader time profile due to an increased number of reflections and scattering inside the TPC. To quantify the detection efficiency, S1 signals are simulated using \texttt {WFsim}. While incorporating a Z-dependent S1 detection efficiency into the detector response model could accurately reflect this aspect, such an approach is computationally intensive. Therefore, the z-dependence is alternatively represented as a systematic uncertainty on the efficiency.

\begin{figure}[h!]
    \centering
    \includegraphics[width=\linewidth]{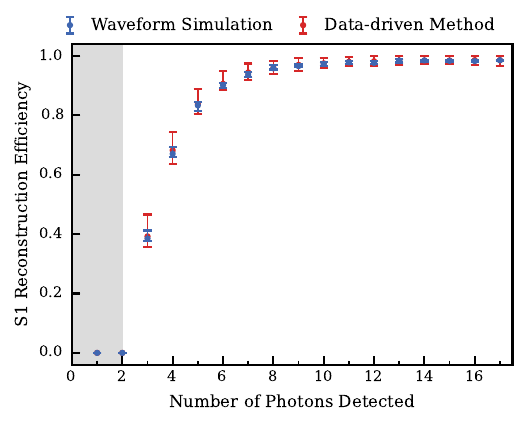}
    \caption{
        S1 reconstruction efficiency as a function of number of detected photons. The red and blue markers show the median detection efficiency for both data-driven and simulation methods. The gray band below 3 photons detected marks the undefined region for S1 when requiring a tight coincidence of at least 3 PMTs. The uncertainty for the data-driven method is mainly a combination of data-selection bias, energy, and position dependence of S1 pulse shape together with statistical uncertainty. The uncertainty for the waveform simulation method is dominated by position dependence in the S1 pulse shape. The final results are based on waveform simulation, while the data-driven method serves as a cross-check.
    }
    \label{fig:s1_effieciency}
\end{figure}

The S1 detection efficiency is shown in Fig.~\ref{fig:s1_effieciency}. The simulation-driven method discussed above was cross-validated using a data-driven method, also shown in Fig.~\ref{fig:s1_effieciency}. In the data-driven method, a subset of photon hit waveforms from larger parent S1 peaks in data are sampled to form smaller S1 signals. These are then processed by the reconstruction software to find S1 peaks. The parent S1 pool is a mixture of S1 peaks from $^{37}$Ar and $^{83\text{m}}$Kr calibration data. The uncertainty band for the data-driven method originates from a combination of data-selection bias, energy and position dependence of the S1 pulse shape and statistical uncertainty. The uncertainty band for the simulation method is dominated by position dependence in the S1 pulse shape.

The S2 efficiency is determined based on a similar simulation-driven procedure. Above 200\;PE, the lowest S2 threshold used, the S2 efficiency is $>99\%$. 

\subsection{Peak Reconstruction Bias}
\label{sec3:peak_rec_bias}
\noindent

The area ($A$) of a peak is the sum of the gain-corrected measured charge in the \pmt{}s, quantified in number of PEs. The reconstruction error $\mathrm{\phi}$ is expressed as the discrepancy between the input area ($A_\mathrm{sim}$) and the output reconstructed area ($A_\mathrm{rec}$),

\begin{equation}
    \mathrm{\phi} = \frac{{A_\mathrm{rec}} - {A_\mathrm{sim}}}{{A_\mathrm{sim}}},
    \label{eq:bias_raw}
\end{equation}

where $A_\mathrm{rec}$ corresponds to the simulated peak area after processing it with the signal reconstruction framework \texttt{straxen}. Several effects contribute to a non-zero value of $\mathrm{\phi}$ for a S1 or S2 peak:
\begin{itemize}
    \item The PMT response to single photon-electron signal can be under-amplified, yielding a negative error in the reconstructed area. 
    \item The per-PMT DAQ digitization threshold prevents very small signals, which might be noise or irrelevant, from being registered. In \texttt{straxen}, the hitfinder threshold works similarly and can result in a negative error.
    \item Electronic noise can distort a signal, which can result in a positive or negative error.
    \item PMT afterpulses and photo-ionization, when merged with their progenitor peaks, will yield a positive error.
    \item The reconstruction software may reconstruct signals too small or too large, for example, if a portion of the signal is wrongfully not considered part of the peak.
\end{itemize}

\begin{figure}[t!]
    \centering
    \includegraphics[width=\linewidth]{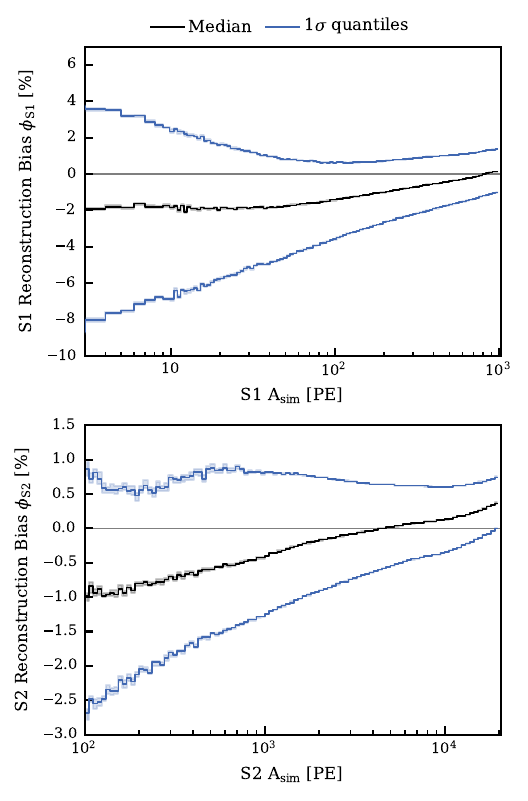}
    \caption{
        Reconstruction bias for S1 (top) and S2 (bottom) signals, calculated according to Eq. \ref{eq:bias_raw}, plotted against the simulated signal area $A_\mathrm{sim}$ with a three-fold tight coincidence requirement. The reconstruction bias corresponding to the median error is shown by the black line, with the $1\sigma$ variation depicted by the blue lines. The shaded bands represent the uncertainties of the quantiles from bootstrapping.
    }
    \label{fig:rec_bias}
\end{figure}
The reconstruction bias is expressed as the median of the $\mathrm{\phi}$ distribution. The bias is estimated by simulating S1 and S2 peaks that are spatially uniformly distributed. The results, showing the median and its $1\upsigma$ quantiles, are presented in Fig.~\ref{fig:rec_bias}.  At low energies, S1 signals exhibit a bias of $-2\%$ due to the digitization threshold and PE under-amplification. This bias exhibits a large spread, as indicated by the $1\sigma$ ranges, which stems from electronic noise and limited statistics. Conversely, at higher energies, an increasing trend in S1 bias is noted, primarily due to the inclusion of afterpulses.
Similarly, for S2 signals, the inclusion of afterpulses results in a positive bias at higher energies. At lower energies ($A_\mathrm{sim} < 10^4$ PE), S2 signals show a negative bias, again influenced by the digitization threshold. 
The bias trends for S2 are otherwise similar to those observed for S1. This peak reconstruction bias study is discussed in more detail in~\cite{Angovaare2023}.

These biases in S1 and S2 signal reconstruction, alongside the S1 reconstruction efficiency, are input parameters for the signal and background response models~\cite{xenonnt_analysis_2} and for the energy reconstruction methodology outlined in Sec.~\ref{sec:7}.

\section{Position reconstruction}
\label{sec:4}

An accurate position reconstruction is crucial for background model building~\cite{xenonnt_analysis_2} and proper signal corrections. The self-shielding effect of LXe keeps radiogenic backgrounds mainly near the edge of the detector, which can be rejected by selecting a restrictive fiducial volume (see Sec.~\ref{sec:7}). Accurate energy reconstructions require position-dependent S1 and S2 corrections (see Sec.~\ref{sec:5}). 

\subsection{3D Position Reconstruction}
\label{subsec3:pos_rec}

The vertical position of an event ($Z_{\text{obs}}$) is obtained by the time difference between its corresponding S1 and S2, namely drift time ($\Delta t_\mathrm{drift}$), multiplied by the expected electron drift velocity, measured in-situ to be \SI{0.675 \pm 0.006}{mm/\us} at 23\;V/cm. The reference depth zero point (Z = 0\;cm) is set at the bottom of the gate electrode, and the maximum drifting distance is set to be $Z = \SI{-148.65}{cm}$, corresponding to the top of the cathode electrode. 

The horizontal position ($\text{X}_{\text{obs}}, \text{Y}_{\text{obs}}$) is obtained from the S2 signal's hit pattern on top PMT array by machine learning based models trained on simulated events. Three independent algorithms were developed using TensorFlow~\cite{tensorflow2015-whitepaper}: one based on a multilayer perceptron (MLP), one on a convolutional neural network (CNN), and a third on a graph constrained network (GCN)~\cite{GCNposrec}. All three algorithms were trained on hit patterns from a full-chain simulation with realistic detector conditions, such as real-time PMT gain values and exclusion of problematic PMTs, as described in Sec.~\ref{subsec1:detect_conditions}. The top light pattern normalized by the maximal PMT signal, which was found to give the best reconstruction performance, is then fed into the models to calculate the horizontal position of an event.

The three algorithms reconstructed most events at the same locations, with differences smaller than a few millimeters. This difference between reconstructed positions is used as an event quality criterion (see Sec.~\ref{sec:6c}). Unless stated otherwise, MLP-based results were used as the default event positions for corrections and analyses since these provided the best resolution.

\subsection{Position Resolution}
\label{subsec3:spatial_resolution}

The position resolution primarily depends on the size of S2 signals or, more specifically, the size of the top PMT's responses to S2 signals, called S2$_\text{top}$. Regions around turned-off PMTs and near the edge of the TPC show worse resolution due to reduced sensors proximity and photon reflections on the PTFE wall.  

\begin{figure}[h!]
  \centering
  \includegraphics[width=\linewidth]{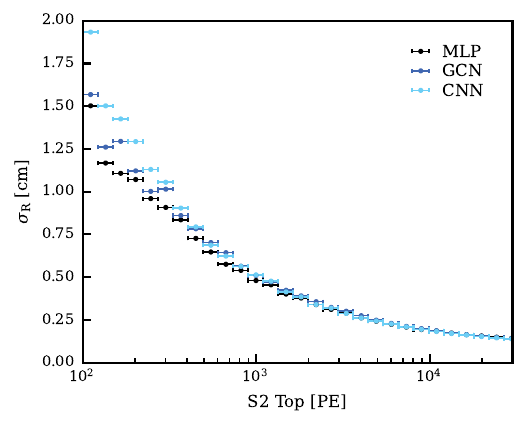}
  \caption{Simulation-driven estimate of the radial resolution in the inner part of the TPC for the three position reconstruction algorithms used in this work within a radius of 60\,cm.}
  \label{fig:posrec:performance}
\end{figure}

With the full chain simulation described in Sec.~\ref{subsec1:detect_conditions}, the position resolution was calculated by comparing the true positions of simulated data and reconstructed positions. This is expressed by the standard deviation $\sigma_\mathrm{R}$ of the differences, shown in Fig.~\ref{fig:posrec:performance} as function of S2$_\text{top}$. Within a radius of 60\,cm, for the lowest energy S2s around 100\,PE, the resolution was estimated to be around 1.5\,cm for GCN and MLP and 1.9\,cm for CNN. In contrast, for large S2 signals ($\ge 10^4$\,PE), the resolution improves to less than 0.25\,cm for all three position reconstruction algorithms. At the edge of the sensitive volume, due to reflections on PTFE adding uncertainties in S2 hit patterns, the event resolution is approximately 1.5 times worse for small S2 signals and 2 times worse for large S2 signals compared to events occurring near the center.

\subsection{Field Distortion Correction}
\label{subsec3:Field_correction}

\begin{figure}[h]
  \centering
  \includegraphics[width=\columnwidth]{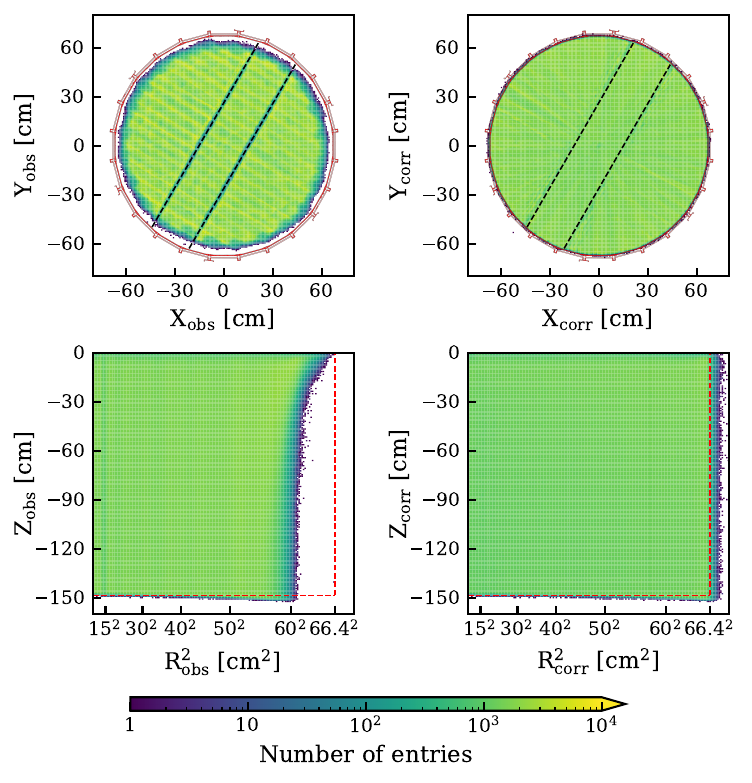}
  \caption{Reconstructed spatial distributions - from MLP algorithms - of $^{83\text{m}}$Kr events in the ($\text{X}, \text{Y}$) space (top) and ($\text{R}^2, \text{Z}$) space (bottom), and pre (left) and post field distortion correction (right). In the ($\text{X}, \text{Y}$) space, the positions of PTFE panels and pillars are illustrated in red, while the black dashed lines indicate the transverse wires installed on the gate electrode. The ($\text{R}^2, \text{Z}$) space includes the cathode position and the PTFE walls in red dashed lines as constraints on the sensitive volume of the TPC.
  }
  \label{fig:sig_cor:FDC_impact}
\end{figure}

Due to the low field, the discreteness of field shaping rings, and negative charge buildup on PTFE walls, the drift field was deformed, resulting in a depth-dependent inward bias of the reconstructed S2 position~\cite{XENON:xenonnt_field_cage}. As shown in the left canvas in Fig.~\ref{fig:sig_cor:FDC_impact}, while $^{83\text{m}}$Kr events should be evenly distributed inside the TPC, events from the bottom were observed more concentrated towards the center as electrons follow the distorted drift field lines. The two distinctive linear features in the ($\text{X}_{\text{obs}}, \text{Y}_{\text{obs}}$) distribution of $^{83\text{m}}$Kr events, shown in Fig.~\ref{fig:sig_cor:FDC_impact}, were caused by the electric-field-channeling effect from additional transverse wires on the gate and anode grids. Moreover, the transversal stride pattern is likely due to the partial shadowing of the top PMT array by anode wires. 

The utilization of wired electrodes added complexities in electric field simulation, making the development of a simulation-driven correction challenging near the transversal wires. To address this, a purely statistical approach, similar to the one used in XENON1T~\cite{Aprile:2019bbb}, was adopted. This method relies on the fact that $^{83\text{m}}$Kr events are uniformly distributed across the radial position within the TPC. Because electric field lines do not cross,
to correct for distortions, each observed radius ($\text{R}_{\text{obs}}$) and its corresponding percentile in the whole population along the radial line can be mapped to an evenly distributed scale from the origin to the TPC's inner wall radius ($\text{R}_{\text{wall}}=66.4$\,cm). The detector was segmented into pie slices based on drift time ($\Delta t_\mathrm{drift}$) and observed azimuthal angle ($\varphi_{\text{obs}}$).  Given that the electric field at the top of the TPC is more disturbed due to its proximity to the electrodes, whereas the field at the bottom is more homogeneous, finer $\Delta t_\mathrm{drift}$ binning was taken at the top. Within each $\Delta t_\mathrm{drift}$ slice, the detector was further divided into sections based on $\varphi_{\text{obs}}$. 

Since no time-dependent PTFE charge-up effects were observed, a single 3-dimensional data-driven field distortion correction (FDC) map, $c(\text{R}_{\text{obs}}, \Delta t_\mathrm{drift}, \varphi_{\text{obs}})$ was created for each of the three machine learning algorithms used for position reconstruction in XENONnT. A detailed description of the FDC map construction can be found in~\cite{Toschi_Francesco_thesis_2022}. The corrected radii ($\text{R}_{\text{corr}}$) for all events are calculated using the following equation: 
\begin{equation}
    \label{eq:fdc_correction_res}
    \text{R}_{\text{corr}} = \text{R}_{\text{obs}} + c(\text{R}_{\text{obs}}, \Delta t_\mathrm{drift}, \varphi_{\text{obs}}).
\end{equation}
The corrected Cartesian coordinates ($\text{X}_{\text{corr}}, \text{Y}_{\text{corr}}$) are derived through trigonometric calculations. The impact of this correction is illustrated in Fig.~\ref{fig:sig_cor:FDC_impact}.
In XENON1T, the $\text{Z}_{\text{obs}}$ position was corrected using a geometric relation that assumes the electron cloud trajectory is a straight line from the interaction point to the extraction point. However, in the case of XENONnT, this assumption does not hold, especially near the very top edge of the TPC. Here, the field distortion is not uniform due to the current configuration of the field shaping rings, and the perpendicular wires can create significant distortions in the electron path~\cite{XENON:xenonnt_field_cage}. Therefore, the field distortion to $\text{Z}_{obs}$ was discarded as it poorly correct the distortion of the cathode at larger radii, while introducing a strong artifact around the top edge of the detector.

\section{Signal Corrections}
\label{sec:5}

The reconstructed S1 and S2 signals have spatial and temporal dependences influenced by detector effects such as electric field inhomogeneities, light absorption, and xenon purity. The detector conditions and signal responses were studied, aiming to understand them and develop corrections to ensure a homogeneous response. 

\begin{figure*}[ht!]
    \centering
    \includegraphics[width=\textwidth]{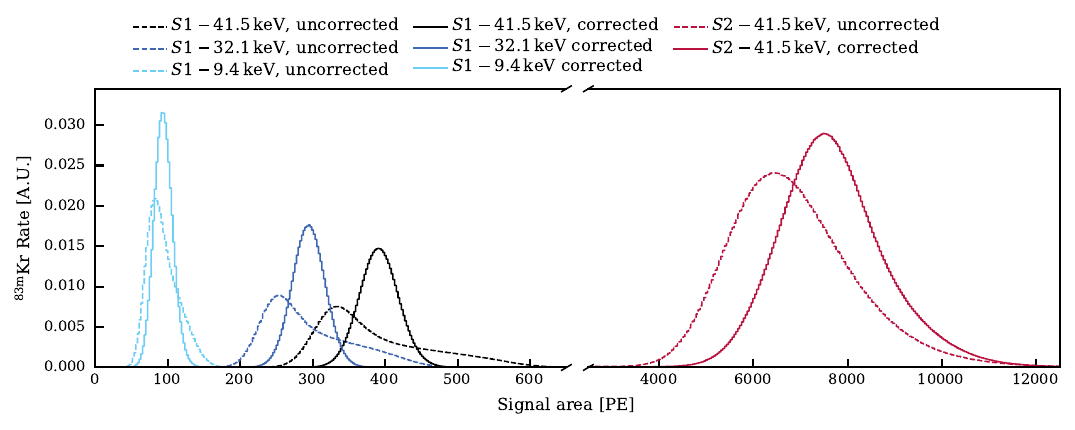}
    \caption{Comparative display of the $^{83\text{m}}$Kr S1 and S2 signals pre- and post-correction in dashed and solid lines respectively.}
    \label{fig:sig_cor:corr_impact}
\end{figure*}

The final impact of the XENONnT SR0 analysis corrections is illustrated using $^{83\text{m}}$Kr calibration data in Fig.~\ref{fig:sig_cor:corr_impact}: an average improvement of $\sim$20\% in signal resolution has been estimated using S1 and S2 signals from the krypton calibration data. In the rest of this section, we will introduce the relevant corrections in detail. 

\subsection{S1 Signal Correction}
\label{subsec4:LCE_cor}

At a given location within the detector in cylindrical coordinates $(R, \varphi, \text{Z})$, the observable photon count is contingent upon several efficiencies: the photon yield (PY) represents the number of photons produced per unit of energy $\varepsilon$ under a specific drift field $E_{\mathrm{drift}}$; the light collection efficiency $\epsilon_L$ indicates the proportion of emitted photons that arrive at a PMT photocathode; the quantum efficiency $\epsilon_{\mathrm{QE}}$ of the PMT is the probability for the incident photon to be converted to a photoelectron by the photocathode; and the PMT collection efficiency $\epsilon_{\mathrm{CE}}$ reflects the efficiency with which PEs are gathered within the PMT itself. These factors collectively determine the LY, i.e., $\langle \mathrm{S1}\rangle$, denoting the number of detected PE per unit of deposited energy in the detection medium:

\begin{equation}
\begin{aligned}
& \mathrm{LY}(\text{R}, \varphi, \text{Z}, \varepsilon) = \frac{\langle\mathrm{S1}\rangle(\text{R}, \varphi, \text{Z}, \varepsilon)}{\varepsilon} \\
& =\epsilon_L(\text{R}, \varphi, \text{Z}) \cdot \mathrm{PY}(\varepsilon, E_{\mathrm{drift}}(\text{R}, \varphi, \text{Z})) \cdot \epsilon_{\mathrm{QE}} \cdot \epsilon_{\mathrm{CE}}.
\end{aligned}
\label{eq:light_yields}
\end{equation}

It should be noted that $\epsilon_{\mathrm{QE}}$ and $\epsilon_{\mathrm{CE}}$ vary by PMTs and thus have spatial dependence. Besides, these factors depend on the physical position where the energy deposition happens. Therefore, the coordinates used in Eq.~\ref{eq:light_yields} are the field-distortion corrected positions as discussed in Sec.~\ref{subsec3:Field_correction}. To enhance readability, we will omit \textit{corr} in subscript for the remainder of this section. The values we used in Eq.~\ref{eq:light_yields} were averaged over all PMTs. The light collection efficiency and field inhomogeneities induce a spatial dependence of the amount of photons collected from the S1 signals. A three-dimensional correction map is derived from the 41.5\;keV signals in $^{83\text{m}}$Kr calibration data to correct for these spatial dependencies. This process necessitates an initial decoupling of effects due to drift field inhomogeneities, which are energy-dependent, from those stemming from the geometric efficiency of light collection, which are not.

As a preliminary step, the $\mathrm{S1}_{\mathrm{fec}}$, representing the field-effect-corrected S1 signals, is calculated. This correction, formulated in Eq.~\ref{eq:S1_field_corr}, involves normalizing the S1 signals by the relative $\mathrm{PY}$ derived from the XENONnT electric field~\citep{XENON:xenonnt_field_cage} and the PY field-dependent model measured in Ref.~\citep{HeXe}. The PY at any given position is adjusted relative to the PY corresponding to the average drift field in the TPC, set at $\langle E_{\mathrm{drift}}\rangle = 22. 9$\;V/cm:

\begin{equation}
\mathrm{S1}_{\mathrm{fec}}(\text{R}, \varphi, \text{Z}) = \mathrm{S1}(\text{R}, \varphi, \text{Z}) \cdot \frac{ \mathrm{PY}(\varepsilon, \langle E_{\mathrm{drift}}\rangle)}{\mathrm{PY}(\varepsilon, E_{\mathrm{drift}}(\text{R}, \varphi, \text{Z}))}.
\label{eq:S1_field_corr}
\end{equation}

To construct the S1 correction map $L_c(\text{R}, \varphi, \text{Z})$ as defined in Eq.~\ref{eq:relative_S1_corr_det}, the TPC is segmented into bins with equal volumes,

\begin{equation}
L_c(\text{R}, \varphi, \text{Z}) = \frac{\mathrm{LY}(\text{R}, \varphi, \text{Z}, \varepsilon, E_{\mathrm{drift}})}{\langle \mathrm{LY}(\varepsilon, E_{\mathrm{drift}})\rangle}  = \frac{\mathrm{S1}_{\mathrm{fec}}(\text{R}, \varphi, \text{Z})}{\langle S1_{\mathrm{fec}} \rangle}.
\label{eq:relative_S1_corr_det}
\end{equation}

\begin{figure}[h!]
  \centering
  \includegraphics[width=\linewidth]{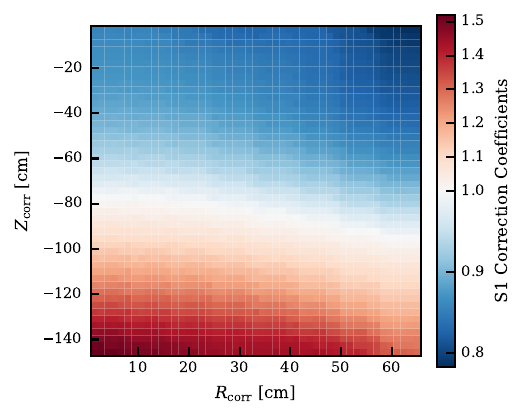}
  \caption{Map of relative light yield extracted from $^{83\text{m}}$Kr calibration data, employed for the S1 correction in SR0. The white region indicates the reference location where the correction factor is 1. And the color map expands linearly above and below. }
  \label{fig:sig_cor:S1_corr_map}
\end{figure}

The number of bins of the correction map is optimized in each dimension to limit the maximum statistical uncertainty to 2\%. For each bin, the average $\mathrm{S1}_{\mathrm{fec}}$ is normalized against the mean $\mathrm{S1}_{\mathrm{fec}}$ observed in the central region of the TPC, defined by the boundaries -130\;cm $<$ $\text{Z}$ $\leq$ 20\;cm and $\text{R}$ $<$ 50\;cm. The resultant map, depicting the relative LY in the detector, is illustrated in Fig.~\ref{fig:sig_cor:S1_corr_map}.

\begin{figure*}[ht!]
  \centering
  \includegraphics[width=0.9\textwidth]{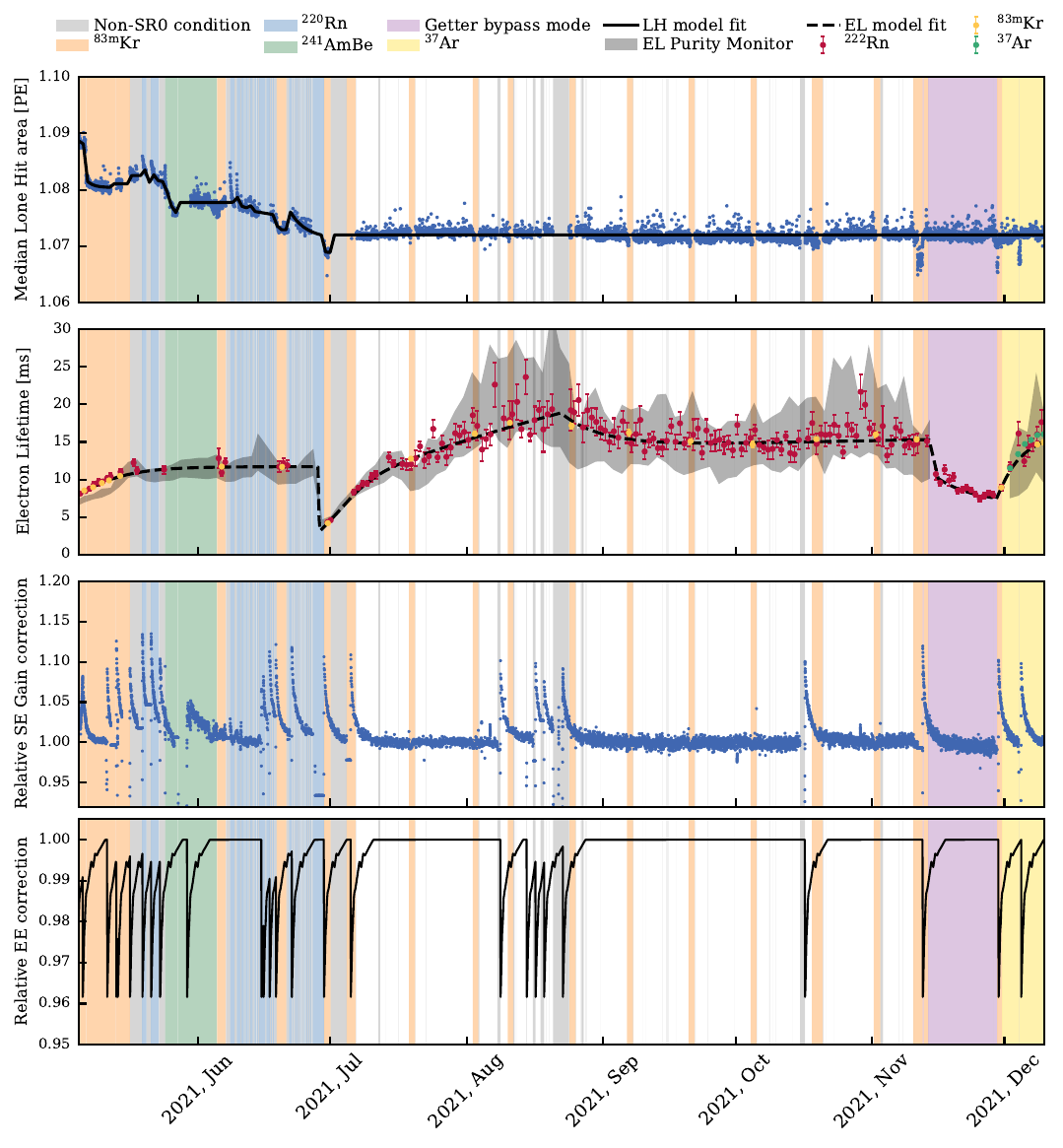}
  \caption{SR0 timeline displaying: (Panel 1) Evolution of median lone-hit area (see Sec.~\ref{subsec:rec_chain}) in top and bottom PMTs, with the time-dependent empirical model used for lone hit correction. (Panel 2) Electron lifetime model derived via $^{222}$Rn (red), $^{83\text{m}}$Kr (yellow), and $^{37}$Ar (green). The gray band is the measurement by the purity monitor with its systematic uncertainty. (Panel 3 and 4) Normalized single electron gain (top) and relative extraction efficiency (bottom) variations, revealing peak structures attributed to the ramping down and up of the anode, induced by occurrences of electron bursts. }
  \label{fig:sig_cor:time_dep_corr}
\end{figure*}

In addition to spatial dependencies, temporal variations in PMT performance influenced the measurements of both S1 and S2 signals. A notable fluctuation in the area of lone hit signals, depicted in the first panel of Fig.~\ref{fig:sig_cor:time_dep_corr}, was observed during the initial calibration phase of SR0. This variation stabilized in the subsequent blinded data-taking period. The evolution of the median lone hit area during the calibration period exhibits two distinct trends. Firstly, a long-term decreasing trend was noted, the exact cause of which remains elusive. However, it is likely attributable to variations in detector conditions, such as temperature and pressure, observed during the same calibration period, which could impact the PMT response. Secondly, short-term fluctuations were detected during calibration periods, coinciding with the PMTs being periodically turned off and on for the injection of \isotope[220]{Rn} sources\footnote{To protect PMTs from the initial high burst of alpha particles in the gas phase after each \isotope[220]{Rn} source injection.}. A relative empirical correction is defined based on the temporal evolution of the median lone hit area for SR0 PMTs, noted $LH_{c}(t)$, normalized against the stable median area observed during the blinded data-taking phase at $LH_0 = 1.07$\;PE. The corrected S1 signal, $\mathrm{cS1}$, after spatial and time-dependent corrections are applied, is computed following Eq.~\ref{eq:cS1}.

\begin{equation}
\mathrm{cS1} = \frac{\mathrm{S1}(\text{R}, \varphi, \text{Z}, t)  \cdot LH_{0}}{L_c(\text{R}, \varphi, \text{Z}) \cdot LH_{c}(t)}.
\label{eq:cS1}
\end{equation}

\subsection{S2 Signal Correction}
\label{subsec4:S2_cor}

To ensure accurate and unbiased energy reconstruction, a given energy deposition should always result in the same S2 area, up to some statistical fluctuations. However, due to detector geometry, PMT responses, imperfect electric fields, etc., the areas of S2 signals show a strong spatial dependence. To remove these detector effects, S2 signals from a mono-energetic $^{\text{83m}}$Kr source were used to generate a correction map to be applied to all observed events and mitigate such instrumental spatial dependence. This correction is referred to as the S2 position-dependent correction.

To protect the PMT arrays from potential damage caused by high-intensity single electron burst events, the gate and anode electrodes were occasionally ramped down when such events occurred during data taking. In close correlation to these ramping-down activities, we observed variations in the S2 yield per electron, called single electron gain (SEG), and a reduction in the fraction of electrons extracted from the liquid to the gas phase, called extraction efficiency (EE), which could be related to the disturbance of the electric field resulting from a relaxation between gravity, electrostatic forces, and wire tensions in the electrodes. Such a dynamic response of the detector is visible up to three days after each ramp-up. Since only relative changes affect the cS2 values, SEG and EE were normalized and implemented as corrections, which we call the S2 time-dependent corrections. 

\subsubsection{S2 Position-Dependent Correction}
\label{subsec4:s2_spatial_cor}

There are two major components for S2 position-dependent corrections: the electron lifetime (EL) correction, which handles the attenuation when electron clouds drift up in the detector, and the S2$(\text{X}, \text{Y})$ correction, which eliminates effects from nonuniform extraction fields due to electrodes sagging or detector tilting and nonuniform light collection. Since the phenomena corrected by S2$(\text{X}, \text{Y})$ all relate to the location of the electron extraction site, the correction depends on the observed location of events, labeled ($\text{X}_{\text{obs}}, \text{Y}_{\text{obs}}$), rather than the inferred ($\text{X}_{\text{corr}}, \text{Y}_{\text{corr}}$) of the initial energy deposition. To enhance readability, we will omit \textit{obs} in subscript for the remainder of this section. 

The electrons inside the drifting cloud can attach to electronegative impurity sites. The resulting S2 attenuation is described by\\

\begin{equation}
\mathrm{S2} = \mathrm{S2}_0 \cdot \exp \left( \frac{-\Delta t_\mathrm{drift}}{\uptau_{\mathrm{EL}}}\right).
\label{eq:s2evolution}
\end{equation}

In this equation, $\uptau_{\mathrm{EL}}$ is the EL, while $\mathrm{S2}_0$ corresponds to the unattenuated S2 area at the interaction vertex depth ($\Delta t_\mathrm{drift} = 0$) before any signal attenuation. The EL is a measure of xenon purity and can be derived from the depth-dependent attenuation of the S2 signal in mono-energetic calibration sources. During SR0, the EL was monitored using $^{83\text{m}}$Kr and \isotope[37]{Ar} calibration data, and 5.6\;MeV $\upalpha$ decays from emanated \isotope[222]{Rn} present in the xenon target. The purity of the xenon target was further evaluated via a purity monitor: a 20\;cm long electron drift chamber in the purification system~\cite{xenonnt_instrumentation}. As for S1 signals, before conducting EL measurements, it is necessary to correct for the drift field's non-uniformity in the TPC.  
The partially corrected S2 as a function of drift time is fitted with Eq.~\ref{eq:s2evolution} to extract the $\uptau_{\mathrm{EL}}$, as demonstrated in~\cite{xenonnt_instrumentation}.

The EL evolution is depicted in the second panel of Fig.~\ref{fig:sig_cor:time_dep_corr}, with the agreement among the three internal sources and the purity monitor. The EL trends during SR0 display multiple sharp declines aligning with the release of impurities during operations on the xenon cryogenics and purification systems\footnote{Such operations include the start of the Rn distillation column, or changes in the xenon purification circuit, which were necessary for the getter bypass mode at the end of SR0.}. Furthermore, continuous outgassing of the materials and restricted xenon circulation flow in the purification systems led to a plateau in the EL after a certain period. The EL measurements in the TPC from $^{83\text{m}}$Kr and \isotope[222]{Rn} data were collectively utilized to model the observed trends, represented by the black dashed line.

\begin{figure}[h!]
    \centering
    \includegraphics[width=\linewidth]{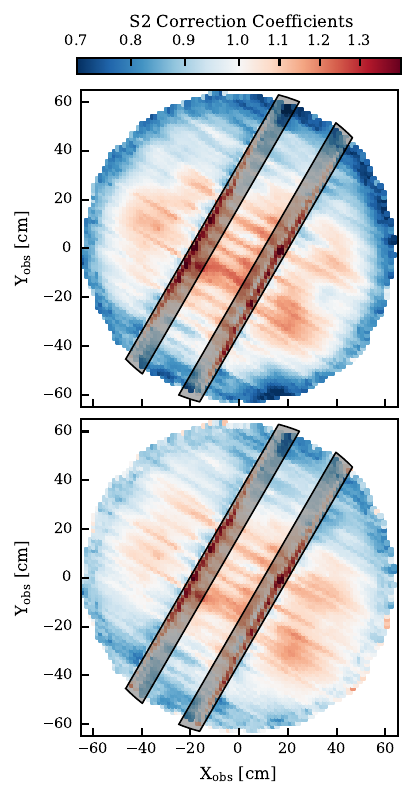}
    \caption{Relative S2 correction maps extracted from $^{83\text{m}}$Kr events, used for S2 correction in XENONnT based on the observed position. The map shown on the top panel has been used to correct the S2 part collected by the top PMT array, whereas the bottom map for the S2 portion was recorded by the bottom PMT array. The white region indicates the reference location where the correction factor is 1. And the color map expands linearly above and below. The near-wire region is shaded out. }
    \label{fig:s2xy_cor_maps}
\end{figure}

To make the S2 signals homogeneous in the XY-plane, the $^{83\text{m}}$Kr calibration data with a total duration of half a month at the beginning of May 2021 were used. These data were divided into runs of 30 minutes. Within each run, the S2s from $^{\text{83m}}$Kr events were normalized to the mean value of the whole population to represent relative S2 area as a function of xy position. Then, the whole (X, Y)-plane ranging from -70\;cm to 70\;cm in both directions were divided into 100$\times$100 bins, within each of which the S2 areas were averaged. This resulted in the expected relative S2 area associated with the ($\text{X}, \text{Y}$) location at the center of the bin. The final S2(X, Y) correction map $c$(X, Y) was then obtained by averaging all the maps with weighted mean, with the number of events from each 30-minute run as weight. 
This method avoids effects from the time evolution of S2 signals due to the electrode operations discussed in Sec.~\ref{subsec4:s2_time_cor} and, thus, the map generated is decoupled from the time evolution corrections. The resulting S2 correction maps to S2$_\text{top}$ and S2$_\text{bottom}$ are shown in Fig.~\ref{fig:s2xy_cor_maps}. A larger correction is required in the center of the TPC, owing to a localized extraction efficiency increase caused by the reduced distance between the electrodes as a result of the sagging of the electrodes. The transversal wires strengthen the extraction field around their locations, resulting in the requirement of exceptionally large ($\sim 50$\%) corrections around the wire regions. Although the PMT responses for S2 signals at the bottom array are more non-localized compared to those at the top array due to the distance to the gas gap, such non-locality only smears away PMT-dependent fluctuations, whereas the absolute S2 yield difference, especially the boost near the transversal wires, is originated from the difference in the number of photons generated by the electrons in the gas gap, and should be observable from the responses of the bottom PMT array. 

While physically decoupled, the developments of S2(X, Y) and EL corrections are correlated and dependent on each other. The method of decoupling the two was to find asymptotic behavior through iteration. Initially, the EL was calculated using uncorrected S2 signals. Following this, a preliminary S2(X, Y) map was constructed. This map was then used to update the EL calculation. This process was iteratively repeated to refine both the map and the EL estimates. Both corrections reach stability after about eight iterations.

\subsubsection{S2 Time-Dependent Correction}
\label{subsec4:s2_time_cor}

\begin{figure}[h!]
  \centering
  \includegraphics[width=\linewidth]{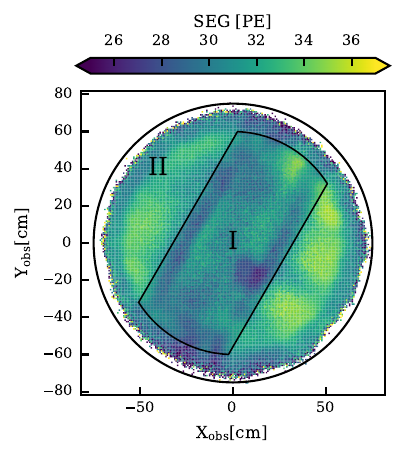}
  \caption{Partitions of the TPC plane due to different single electron gain and relative extraction efficiency responses after each ramp-up of electrodes, overlaid with the SEG X-Y distribution. The SEG in the outer region of the TPC (partition II) is larger due to the increased distance between the liquid-gas interface and the anode. In the near-wire region, the SEG is smaller. Due to the weaker position resolution, the SEG X-Y distribution does not exhibit the same pattern as shown in Fig.~\ref{fig:s2xy_cor_maps}. The black circle is slightly larger than the TPC radius (i.e., 66.4\;cm) because the single electron gain selection accounts for the position resolution, as discussed in Sec.~\ref{subsec3:spatial_resolution}.}
  \label{fig:sig_cor:TPC_coarse_partitioning}
\end{figure}

The S2 time-dependent corrections have two components: the evolutions of SEG and EE. SEG can be obtained by tracking the S2s of the single electron population. Two distinctive behaviors of SEG were observed inside the TPC: between the gate transverse wires, there was a strong change in SEG after ramp up, while outside it was stable. Such a behavior difference led to a partitioning inside the TPC as shown in Fig.~\ref{fig:sig_cor:TPC_coarse_partitioning}. 
To normalize the SEG values to correction factors, SEG values between 2021-07-05 and 2021-08-08 were used, and two SEG values, respectively, in the two partitions, were set to be the reference SEG for SR0. The normalized SEG evolution for partition I (P-I) is shown in Fig.~\ref{fig:sig_cor:time_dep_corr}. No time variation was observed in partition II (P-II), so the correction factors were set to one (no correction) and, therefore, it is not displayed.

A relative extraction efficiency (REE) was calculated based on position-dependence corrected (both S2(X, Y) and EL) S2 signals of $^{83\text{m}}$Kr events and the SEG for each run. By dividing SEG from position-dependence corrected S2 signals of $^{83\text{m}}$Kr events, the expected number of electrons can be obtained from this mono-energetic source. The mean value of the expected number of electrons between 2021-07-05 and 2021-08-08 was used as a reference, and all other values were normalized to obtain REE. These were used to construct a model, as shown in the bottom panel of Fig.~\ref{fig:sig_cor:time_dep_corr} for P-I, which was propagated to the whole SR0. No REE evolution was observed in P-II, therefore the correction factors in P-II were set to one. 

These two time-dependent corrections for S2 signals helped restore background data near the ramping electrode events, representing 7.1\% of the total exposure. The corrected S2 signal, cS2, after spatial and time-dependent corrections are applied, is computed as:

\section{Selection Criteria}
\label{sec:6}

This section describes the criteria applied to the nuclear recoil WIMP search and the low-energy ER analysis~\cite{xenon_nt_wimp.131.041003,xenonnt_lower}. The aim is to select single scatter events in the region of interest while rejecting unphysical events and improperly reconstructed events. 

\subsection{Operating Conditions of Data Acquisition}

A series of selections based on the operational conditions during data acquisition have been used to remove data during certain time periods.

For most science data taking, the DAQ systems of the three detectors were synchronized~\cite{XENON:2022vye}. Synchronization checks among the XENONnT detectors were conducted using a GPS clock signal (0.1\;Hz)~\cite{DeDeo:2019cih}. Initially, during the start of SR0, the GPS clock was not fully operational, resulting in the linked data-taking mode being unavailable for the three detectors for approximately 8\% of the total data collection period. Any periods of linked data acquisition that might have exhibited synchronization loss were to be excluded; however, such instances did not occur during SR0.

If a digitizer's buffer is full, its board cannot accept further data from the photosensors, resulting in partially acquired events. The DAQ veto selection rejects these data acquisition periods, resulting in a 0.04\% reduction of the live time in SR0.

The muon veto detector aims to detect muons and muon-induced backgrounds, particularly fast neutrons from muon spallation and electromagnetic or hadronic muon cascades. Whenever this happens, the TPC data acquisition is vetoed for 1\;ms. This hardware trigger requires signals larger than 1\;PE in at least 10 MV PMTs within a 300\;ns time window. The tagging efficiency for backgrounds induced by muons crossing the water tank, or external muons with a shower in the water tank, is equal to 100\% and 38\%, respectively, as it was estimated in~\cite{Aprile_2014}. The muon veto criterion reduces these backgrounds by a factor of 2.8 with a 1\% loss of livetime.

The neutron veto, operating with pure water in SR0, aims to tag neutrons with one energy deposition in the TPC. These neutrons are detected by observing the Cherenkov light emitted when they are captured by a hydrogen atom. For the NR WIMP search, the NV issues a veto signal for each NV event with at least 5 PMT signals recorded and a total area of at least 5 PE~\cite{xenonnt_nv}. A neutron tagging efficiency of (53 $\pm$ 3)\% is estimated using neutrons from the \isotope[241]{AmBe} source, in coincidence with 4.4\;MeV gammas from de-excitation of $^{12}$C, which originate by the capture of $\upalpha$-particles on $^{9}$Be, recorded by the TPC~\cite{xenonnt_nv}.  This is slightly reduced to (50 $\pm$ 3)\% due to the unlinked data-taking period. For details on the background modelling see~\cite{xenonnt_analysis_2}. Motivated by the estimated characteristic neutron-capture time of (174 $\pm$ 11)\;$\upmu$s, all the S1 signals in a time window (-1, 249)\;$\upmu$s around the center time of the neutron-veto event are vetoed~\cite{xenon_nt_wimp.131.041003}.  

Conversely, the ER search uses the coincidence of 3 NV PMTs as a trigger requirement. The lower threshold is acceptable as the veto window is reduced to a $\pm$300\;ns window in which S1 signals of a TPC event are vetoed~\cite{xenonnt_lower}. A factor of 8\% reduction in the gamma-ray contribution from the decay of radionuclides in the detector material is estimated from science data. 

The livetime loss due to NV selection is 1.6\% and 0.03\% for the WIMP and low-energy ER analyses, respectively.

\subsection{Accidental Coincidence Suppression}

Accidental coincidences (AC) arise from the random pairing of isolated S1 and S2 peaks. A few plausible origins may be fake S1 signals from pileup lone PMT hits, misclassified single electrons mistaken for S1 signals, and S2s from pileup few-electron signals arising from various sources.

A series of selections based on the temporal and spatial correlation of these events with previous large peaks are used to suppress AC-like events. Events occurring within one maximum drift time from an S1 (S2) signal larger than 10$^3$\;PE (10$^4$\;PE) are vetoed. For a time difference $\Delta t_\mathrm{prev}$ larger than one maximum drift time, S2 peaks occurring too close to a triggering S2 peak are vetoed based on a quantity called ``time shadow'' defined as:

\begin{equation}
\mathrm{S2}_{\text{time shadow}} = \mathrm{S2}_\text{prev} \mathbin{/} \Delta t_\mathrm{prev}.
\label{eq:time_shadow}
\end{equation}

\begin{figure}[t!]
    \centering
    \includegraphics[width=\linewidth]{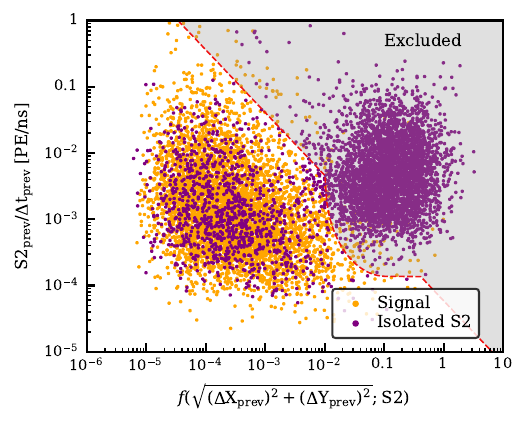}
    \caption{S2 Position Shadow Distribution. The time shadow quantity is shown as a function of the Half-Cauchy distribution of position differences between an isolated S2 signals and its preceding S2 peak, with the X-scale reflecting the peak's position reconstruction accuracy. The S2 signals of a true event and previous S2s have no positional correlation and accumulate in the left region. However, for isolated S2 signals, the opposite is true. This can be used to distinguish between signal and isolated peaks.
    }
    \label{fig:position_shadow}
\end{figure}

A threshold of 0.038\;PE/ns allows suppression of isolated S1s peaks while maintaining 96\% signal acceptance. The spatial distance between previous large S2 signals and isolated S2s, described as a Half-Cauchy distribution $\rho(\sqrt{\Delta \text{X}^2 + \Delta \text{Y}^2})$~\cite{Aprile:2019bbb}, is used to reduce the AC-like events due to delayed extracted electrons. The rejection region is a function of S2$_{\text{time shadow}}$, as shown in Fig.~\ref{fig:position_shadow}. This selection retains $\sim$97\% signal acceptance.

The impact of a large S2 signal on the single electron (SE) rate is significant, resulting in an increase in both the misclassification of S1 signals as SE (leading to isolated S1 signals) and the occurrence of SE pile-up (leading to isolated S2 signals). AC-like (S1, S2) pairs are suppressed based on the number of peaks occurring in a 2\;ms window before the S1 and within a radius of 6.7\;cm from the S2 signal. The threshold of the number of peaks allowed is chosen to satisfy $\sim$99\% signal acceptance.

To further mitigate AC-like events in the WIMP analysis, a Gradient Boosted Decision Tree (GBDT) algorithm~\cite{scikit-learn}, is employed. This algorithm is trained using simulated signal events to enhance its effectiveness. Five features are used in training: S2 rise time (see Sec.~\ref{subsec:rec_chain}), the time interval in which 50\% and 90\% quantiles of the S2 peak are contained, the S2 peak area, and the observed Z position. The algorithm returns a metric, or score, for each event to be signal-like and AC-like. To avoid loss in the signal acceptance, two algorithms are used depending on whether the S2 area is smaller or larger than 2000\;PE. The rejection thresholds for the GBDT scores, equal to 0.84 and 0.55, respectively, have been chosen to guarantee an acceptance of more than 95\% in the region of interest.

\subsection{Signal Reconstruction Requirements}
\label{sec:6c}
\begin{figure}[t!]
    \centering
    \includegraphics[width=\linewidth]{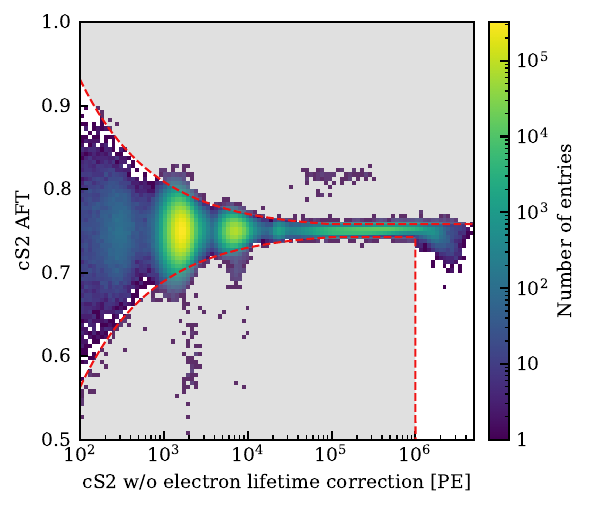}
    \caption{Distribution of cS2 area fraction top vs. cS2 signal area without electron lifetime correction as measured in \isotope[241]{AmBe} and $^{37}$Ar calibration data. The dashed red line demarcates the selection criteria.}
    \label{fig:cs2_aft_cut}
\end{figure}

Incorrectly reconstructed events, members of known background populations, and low-quality signals are removed by a set of requirements on the reconstructed signals.
\begin{itemize}
    \item Large deviations in the results of the different position reconstruction algorithms indicate a modeling error or an abnormal event. Data quality is improved by rejecting events with a position difference greater than the 99\% quantile of position differences seen in high-quality calibration data (see Sec.~\ref{subsec1:Sciences_runs}). The quantile is a function of S2 area because the position reconstruction performance varies with the S2 area.
    \item Events with S1s dominated by one PMT are suspicious. Typically, they are caused by a PMT malfunction, such as afterpulses or light emission~\cite{PMT_XENONnT:Antochi2021}. At the top of the detector, events where a PMT contributes more than 6\% plus an offset of 4\;PE to the S1 are rejected. Events deeper in the detector have more concentrated S1 hit patterns, so the rejection boundary is linearly increasing to 13\% for events reconstructed at the cathode. 
    \item Events with an S1 characterized by abnormal temporal widths are typically due to misidentified SE signals. They are excluded by special selection criteria based on the 50\% and 90\% quantiles of the S1 peak. The rejection regions are defined based on the 99\% quantiles in the examined parameter spaces as a function of the signal size observed in high-quality calibration data.
    \item Events whose S2 is either a misidentified S1 or caused by gas phase interactions above the anode electrode are removed using the fraction of the S2 signals light collected by the top PMT array. Events with a corrected S2 light fraction in the top array outside the 0.5\% and 99.5\% quantiles, defined as a function of the corrected S2 area, are discarded. As before, the quantile lines are derived from a high-quality calibration data sample and are used as rejection limits up to 10$^5$\;PE; above this, a constant extrapolation is used. To avoid unnecessary loss of signal acceptance, the lower boundary is removed for signals larger than 10$^6$\;PE. The S2 top array light fraction is only related to the light transmission, so the selection is based on the S2 signal before the electron lifetime correction. The rejection criterion is shown in Fig.~\ref{fig:cs2_aft_cut}.
    \item The consistency between the observed and expected S2 pattern intensity distribution on the top PMT array is quantified by a $\chi^2$ goodness-of-fit test and outlier events are rejected. The expected light distribution is derived from a data-driven map built on a neural network fed with high quality $^{83\text{m}}$Kr calibration data. The 99\% quantile of the $\chi^2$ distribution as a function of S2 area is used as the selection criterion. This selection mainly suppresses the pileup of delayed electron signals, double scatters, and mis-reconstructed events. Due to a loss of accuracy in the data-driven map for large S2 areas, this criterion is limited for events with S2s smaller than 3$\times$10$^4$\;PE.
    \item A Naive Bayes Classification (NBC) method~\cite{XENONCollaboration:2023dar} is used to quantify whenever the S1 and S2 have the expected waveform shape. This machine learning algorithm, based on a 50-sample waveform of the peaks and on a 50-sample quantile representation of peak waveforms, assigns each peak a score indicating the accuracy with which they were reconstructed. Events in the 99\% quantile line in the parameter space of S1 (S2) score and S1 (S2) size are not further considered in the analyses. These selections effectively remove misclassified single electrons, gas events, unresolved double scatters, and after-pulse contaminated S1s from the data set.
\end{itemize}

\subsection{Event Requirements}
A high quality of S1-S2 pairing is guaranteed by selection criteria developed by exploiting the correlation of event features and the position of the original interaction.
\begin{itemize}
    \item Events with anomalous S1 light pattern distributions, e.g., from unresolved multiple scatters or partially reconstructed events, can be rejected by comparing the S1 hit patterns with the expected patterns derived from optical MC simulations. 
    The selection criteria are tuned to accept S1-S2 signal pairs from physical interactions in calibration control samples with a probability greater than 99\%.
    \item Similarly, the correlation between the fraction of the S1 signal observed by the top array and the reconstructed event position is used to reject unphysical events. This quantity follows a binomial process: each observed photon is either seen by the top or bottom PMT array, and the probability depends solely on the event's location and detector geometry. A well-motivated data quality criterion, based on the p-value of the binomial test, is used to suppress accidental S1-S2 pairings and poorly reconstructed events.

    \item The ionization electron cloud, created by a particle interaction in liquid xenon, diffuses over time, which affects the features of the reconstructed S2s. The S2 width, or the time interval $r_{50}$ in which the 50\% quantile of the S2's area is contained, is correlated with the drift time $\Delta t_\mathrm{drift}$ of the event. Diffusion ensures a Gaussian distribution for the electron cloud, so this can be described~\cite{Sorensen_2011} as:
    \begin{equation}
        r_{50} = \sigma_{50} \sqrt{ \frac{2D_L \cdot (\Delta t_{\text{drift}} - \Delta t_\mathrm{gate})}{v_d^2}},
        \label{eq:r_50}
    \end{equation}
    where $\sigma_{50} \approx 1.35$ is the conversion from Gaussian standard deviation to the 50$^{\text{th}}$ area range, $D_L$ is the field-dependent longitudinal diffusion coefficient, and $v_d$ is the drift velocity. For the purpose of modeling solely the electron diffusion, the observed drift time must be corrected for the drift within the extraction field, from the gate to the liquid-gas interface, $\Delta t_\mathrm{gate}$. The three parameters required to model the S2 width  are drift field dependent and are determined from $^{83\text{m}}$Kr calibration data: $D_{L} = (45.6 \pm 0.1)\; \text{cm}^2~\text{s}^{-1}$, $v_{d} = (0.675 \pm 0.006)\;\text{mm}~\upmu\text{s}^{-1}$, and $\Delta t_\mathrm{gate} = (4.4 \pm 0.5)\;\upmu \text{s}$. While the expected value of the S2 width of an event depends only on the drift time, the spread of the distribution of S2 width is highly dependent on the size of the S2 signal. Therefore, an S2 area-dependent selection based on the ratio between the observed width of an event and the width expected from a model, called the ``normalized width'' $r_{50}^{\text{norm}}$, is used to ensure correlation: 
    \begin{equation}
        r_{50}^{\text{norm}}(\Delta t_\mathrm{drift}) = \frac{r_{50}(\Delta t_\mathrm{drift})^2 - (r_{50}^{SE})^2}{r_{50}^{\text{model}}(\Delta t_\mathrm{drift})^2}.
        \label{eq:r_50_norm}
    \end{equation}
    The quantity $r_{50}^{SE} \sim 375\;\text{ns}$ is introduced to correct the width model for a small number of electrons in the electron cloud and is calculated from the observed single electron population. For events under the perpendicular wires, the observed S2 widths are larger than in the rest of the TPC due to longer drift times caused by the lower field region right below the wires. Thus, the selection follows a different definition for events reconstructed within 4.45\;cm of the perpendicular wires. In Fig.~\ref{fig:cut_far_wires}, the boundaries of the cut in the region far from the wires are shown. They are defined as the 1$^{\text{st}}$ and 99$^{\text{th}}$ percentiles of the distributions of signal-like simulated data for S2 signal less than 10$^{4}$\;PE and otherwise on $^{220}$Rn data. The broadening of the distribution for small S2 signal sizes is caused by binomial fluctuations in the number of electrons contributing to the signal. 
    \begin{figure}[t!]
    \centering
    \includegraphics[width=\linewidth]{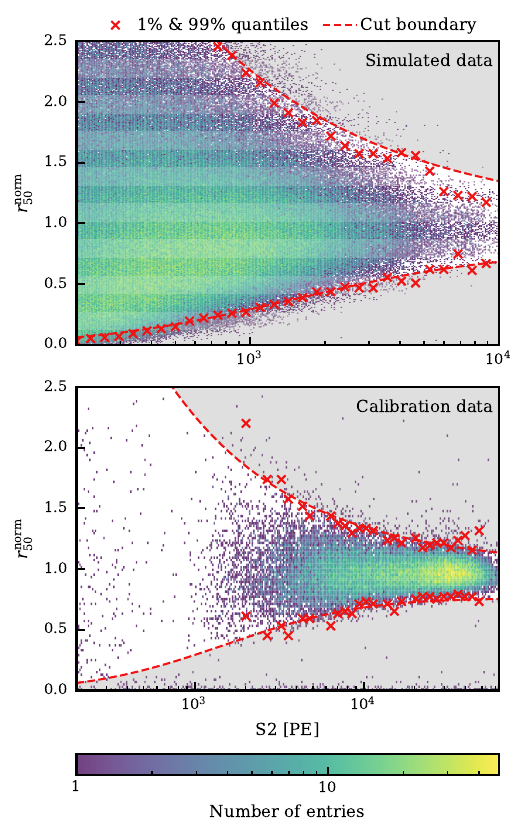}
\caption{Distribution of the measured S2 width parameter r$_{50}^{\text{norm}}$ as a function of S2 signal area for simulated (top) and $^{220}$Rn calibration data (bottom) for events reconstructed far from the perpendicular wires. Red crosses mark the 1\% and 99\% quantiles, and the dashed line shows the selection criterion definition for the far-wire region.
    }
    \label{fig:cut_far_wires}
\end{figure}    
    In the near wire region, the 1$^{\text{st}}$ and 5$^{\text{th}}$ percentiles in the (S2, $r_{50}^{\text{norm}}(\Delta t_\mathrm{drift})$) parameter space of the observed data are used as the rejection limit for the low-energy ER and WIMP searches, respectively. The different boundary conditions were motivated by varying AC contributions for the two analyses. For the low-energy ER analysis, we preferred higher signal acceptance, whereas for the WIMP search, we chose to optimize AC background suppression. In addition, the difference between the $r_{50}(\Delta t_\mathrm{drift})$ and $r_{50}^{\text{model}}$ as a function of the distance from the perpendicular wires is used to discriminate anomalous events. The selection definition is portrayed in Fig.~\ref{fig:cut_near_wires}.
\begin{figure}[t!]
    \centering
    \includegraphics[width=\linewidth]{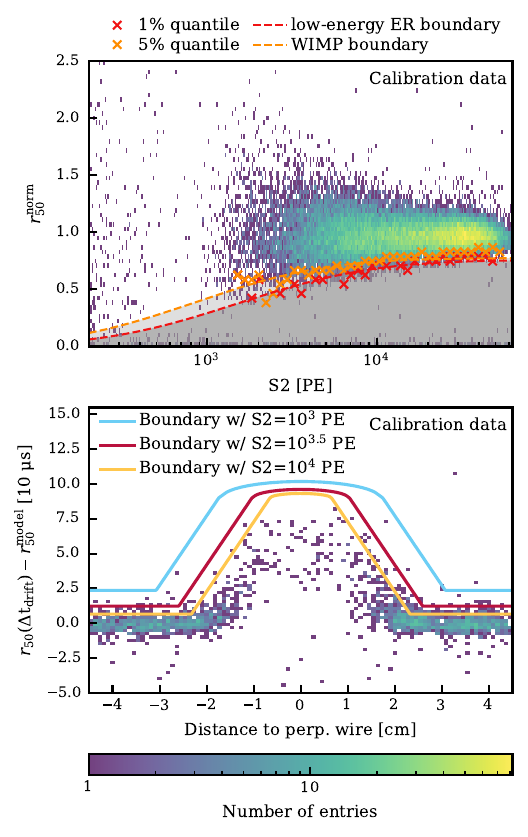}
    \caption{Distribution of the measured S2 width parameter r$_{50}^{\text{norm}}$ as a function of S2 signal area for $^{220}$Rn calibration data reconstructed in the near-wire region. The orange and red crosses show the 1\% and 5\% quantiles of the data, which are used to define the selection criteria used in the main XENONnT analysis, as depicted by the legend (top). Difference between expected and measured S2 width parameter r$_{50}$ as a function of the distance of the reconstructed position to the perpendicular wires for $^{220}$Rn calibration data. Three lines are shown: the S2 width cut upper boundaries for near-wire regions for different S2 signal areas (bottom).
    }
    \label{fig:cut_near_wires}
\end{figure}
\end{itemize}

\subsection{Single-Scatter Requirements}

\begin{figure}[t!]
    \centering
    \includegraphics[width=\linewidth]{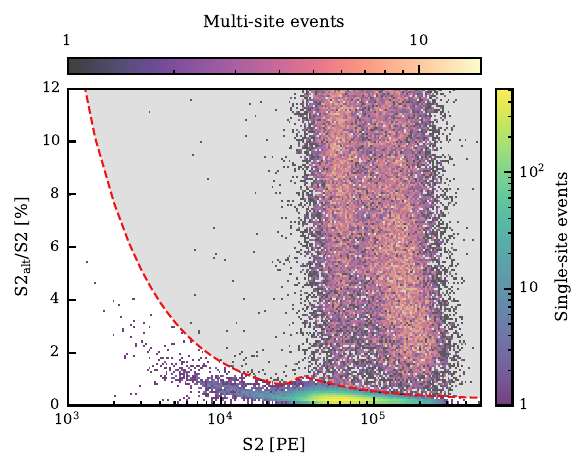}
    \caption{Distribution of S2$_{\text{alt}}$/S2 as a function of S2 area for $^{220}$Rn calibration data. The dashed line defines the adopted S2 single scatter selection definition and divides the single-site and the multi-site populations. Approaching high S2 signals, the selection definition is relaxed due to the non-optimal \texttt{straxen} performance for high energy peak splitting/merging. The two populations in the multi-site events are due to the gamma transitions followed by the beta decay of $^{212}$Pb, daughter of $^{220}$Rn.
    }
    \label{fig:s2_ss_cut}
\end{figure}

Given the small expected scattering cross-section of dark matter particles and the small mean free path of photons and electrons in the energy range of interest, the signals searched (WIMP and other low-energy ER signals) are expected to have only single energy depositions in the TPC. Identifying multiple scatter events is a powerful discriminator between signal candidates and certain backgrounds. For example, radiogenic neutrons have a probability larger than 80\% to induce multi-site events. 
\begin{itemize}
    \item Events with alternative S1s recorded in the waveform that could also form a valid interaction with the primary S2 are rejected. Whether the alternative S1 and the main S2 do not constitute a valid interaction is based on the S2 width, S1 AFT, and the light distribution of the alternative S1. S1 signals with abnormally high contributions of a single PMT are not considered for pairing with the S2. This selection not only targets multi-energy deposition events but also events with ambiguous identification of the primary S1.
    \item Events with an additional S2 in the waveform are considered multiple scatter if the signal ratio with the main S2 is larger than a few percent. The threshold is based on high-quality calibration data and is defined as a function of the primary S2 area, as shown in Fig.~\ref{fig:s2_ss_cut}. As for S1 signals, events in which every alternative S2 is an unphysical artifact are valid single scatter events.
\end{itemize}

\subsection{Fiducial Volume}

The rejection of the periphery of the detector is the most robust selection against poorly reconstructed events and several backgrounds. 

The fiducial volume (FV) optimization uses the background components’ expected (R$^2$, z) distributions, excluding regions where detector understanding is limited.
The optimization region considered falls below 100\;PE S1 and loosely within either the NR or ER bands. Specifically, it is between the 1$^{\text{st}}$ percentile of a 50\;GeV/c$^2$ WIMP in cS2 and the 99$^{\text{th}}$ percentile of the ER background (or a low-energy ER signal) in cS2.
ER background is mostly from homogeneously distributed \isotope[214]{Pb} $\upbeta$-decays and inhomogeneously distributed $\upgamma$ emission from detector materials. Their position distribution is modeled using unblinded background events with reconstructed energy between 20\;keV and 40\;keV. The AC model is data-driven, constructed using unpaired S1s and S2s from the physics data, randomly paired to build high-statistics artificial data, and validated against calibration data. The spatial distribution of ACs is approximately constant over the detector volume and, thus, does not substantially impact the fiducial volume optimization. Events near the TPC wall, originating, e.g., from the $^{222}$Rn progeny plated out on the inner surface of the PTFE panel, tend to lose a fraction of their charge in the PTFE panels~\cite{xenonnt_analysis_2}. This leads to events characterized by a low charge-to-light ratio, which may be inaccurately reconstructed further inside the TPC radius. These events are modeled using sidebands of blinded WIMP search data~\cite{xenonnt_analysis_2}. NR events are expected from radiogenic neutrons produced through spontaneous fission or ($\upalpha$, n) reactions in detector materials and from coherent elastic neutrino-nucleus scattering from neutrino of astrophysical origin~\cite{xenonnt_cevns_2024}. The spatial distribution of the latter is uniform; therefore, it is not considered during the optimization of the fiducial volume. For the background of radiogenic neutrons, the neutron yield is simulated using \texttt{Geant4}, and it is adjusted to match the expected number of background events as forecasted by the SOURCES-4A simulation package~\cite{osti_15215, TOMASELLO2008431}. The derived position distribution has been propagated to determine the choice of fiducial volume.

\begin{figure}[h!]
  \centering
  \includegraphics[width=\linewidth]{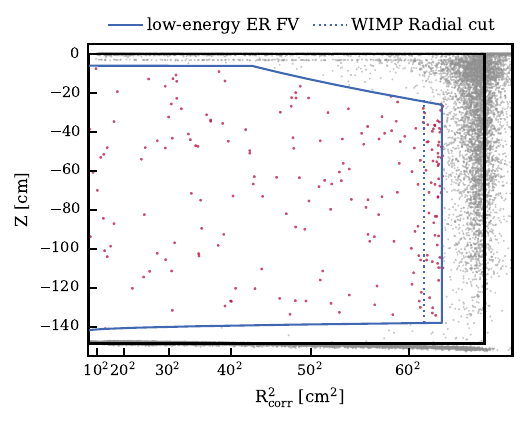}
  \caption{Observed data in (R$^2$, Z) space of the SR0 WIMP search. The red dots are the events that have been reconstructed in the fiducial volumes. The solid black box demarcates the TPC volume, whereas the solid and dotted blue lines show the adopted fiducialization for the main XENONnT SR0 analyses~\cite{xenonnt_lower, xenon_nt_wimp.131.041003}.}
  \label{fig:cut:fiducial_volume}
\end{figure}

To exclude mis-reconstructed events from the gas volume, events reconstructed with $\text{Z} < -6$\;cm are excluded. Additionally, TPC regions where the difference between simulation- and the data-driven electric field is larger than 10\% are not further considered. These are well confined at a high radius and close to the cathode and gate electrodes. The top right corner is also removed due to the high ER background and relatively high electric field variation in that region. Lastly, the maximum radius is set as 63\;cm for the low-energy ER analysis and 61.35\;cm for the WIMP search to reject the bulk of surface background events, as shown by the solid and dashed blue lines in Fig.~\ref{fig:cut:fiducial_volume} respectively.

The xenon mass contained in the FV is derived from geometrical considerations, assuming a liquid xenon density of (2.862 $\pm$ 0.003)\;t/m$^3$~\cite{doi:10.1021/ie4033999}, considering the presence of S2-insensitive mass (see \cite{XENON:xenonnt_field_cage} for additional information) and given the best knowledge of the drift field as well as the resolution of the position reconstruction. The field distortion correction defined for SR0 does not include the effect of a small charge-insensitive volume near the cathode and at the periphery of the TPC, which effectively reduces the maximum radius of the TPC. We do consider the effect that this has on the mass contained in the fiducial volume. The FVs contain (4.37 $\pm$ 0.14)\;t and (4.18 $\pm$ 0.13)\;t of liquid xenon for the low-energy ER analysis and WIMP search, respectively. The uncertainties include the position reconstruction resolution (0.1\%) and the dimension of the charge insensitive mass based on electric field simulations (3\%).

\subsection{Signal Acceptances}

\begin{figure*}[]  
  \centering
  \includegraphics[width=\textwidth]{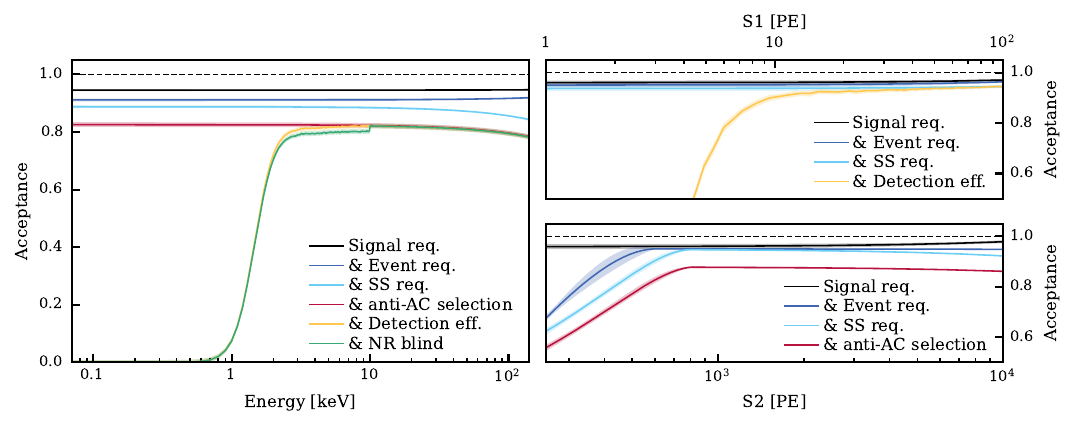}
  \caption{Evolution of the acceptance by incrementally applying the selection criteria categories described in the text as a function of reconstructed energy for low-energy ER analysis (left) and uncorrected S1 and S2 signal areas for WIMP search (right). The shade bands represent the acceptance uncertainty. The 10\;keV discontinuity in total acceptance as a function reconstructed energy marks the WIMP-blinded region, relevant only for low-energy ER analysis. The total acceptance is also shown after considering the reconstruction efficiencies.}
  \label{fig:sr0_wimp_lower_acc}
\end{figure*}

Fig.~\ref{fig:sr0_wimp_lower_acc} shows the cumulative signal acceptances of the described categories of selection criteria as a function of the reconstructed energy and the uncorrected S1 and S2 signal sizes. The acceptance of each selection, namely the signal events that pass through the selection, is estimated by using the \textit{N-1 method}, namely evaluating the  N$^{\text{th}}$ selection acceptance after applying the previous N-1 criteria. The uncertainties in the acceptances were inferred using the Clopper-Pearson method~\cite{10.1093/biomet/26.4.404}. The acceptances were estimated using $^{220}$Rn and $^{37}$Ar calibration data in a $\sim4$ tonne fiducial volume, equivalent to R $<$ 60.73\;cm and Z $\in[-13.6; -134.2]$\;cm. Complementary,  synthetic signal-like events from the waveform simulation were used to establish the acceptances, e.g., for the accidental coincidence and S2 width criteria. Whenever a significant correlation is observed between two or more selections, these are grouped, and their cumulative acceptance is estimated. This is the case for S2 width and GBDT anti-AC criterion. Selections dealing with properties unrelated to the event are deemed exposure reduction cuts, e.g., fiducialization or selections based on operational conditions. The smooth curves in Fig.~\ref{fig:sr0_wimp_lower_acc} are determined by fitting polynomial functions to the data points. The uncertainty bands account for the uncertainty from the fitting procedure. 

In the S1 signal space, the selection criteria have a similar impact with an average acceptance equal to (98.1 $\pm$ 0.9)\%. In the S2 signal space, the acceptance is primarily influenced by the anti-AC requirements and the event quality criteria, particularly the S2 width selection criterion. The discontinuity at 10\;keV in the total acceptance as a function of the reconstructed energy accounts for the WIMP blinded region, and it is relevant only for low-energy ER analysis. After including the reconstruction efficiencies discussed in Sec.~\ref{subsec:rec_eff}, the total acceptances are propagated into the statistical inference for dark matter and low-energy ER searches.

\section{Energy reconstruction and resolution}
\label{sec:7}

The energy deposited in an ER interaction ($E_{\mathrm{ER}}$), which is converted into scintillation photons $n_{\mathrm{ph}}$ and ionization electrons $n_{\mathrm{e}} $, can be expressed as a function of the reconstructed cS1 and cS2 signals by introducing the photon detection efficiency $g_1$, also known as photon gain, and electron gain $g_2$:

\begin{equation}
E_{\mathrm{ER}} = (n_{\mathrm{ph}} + n_{\mathrm{e}}) \cdot W = \left(\frac{\mathrm{cS} 1}{g_1} + \frac{\mathrm{cS} 2}{g_2}\right) \cdot W,
\label{eq:energy_transformation}
\end{equation}

where $W = (13.7 \pm 0.2)$\;eV/quantum~\cite{Dahl:2009nta} represents the mean energy required to create either scintillation or ionization quanta. The $g_1$ and $g_2$ factors are detector-dependent parameters assessed using mono-energetic peaks, including  \isotope[37]{Ar}, \isotope[83\text{m}]{Kr}, \isotope[129\text{m}]{Xe}, and \isotope[131\text{m}]{Xe}. Higher energy lines of \isotope[60]{Co}, \isotope[40]{K}, \isotope[214]{Bi}, \isotope[12]{C}, and \isotope[2]{H}, are excluded from the fit due to missing high-energy optimizations of the signal reconstruction and correction, but still reported for completeness. The selection criteria outlined earlier are applied to all data, with the exception of the \(^{83\text{m}}\)Kr calibration, which uses dedicated topology-based cuts. The measurement for each source of the mean charge yields ($\mathrm{CY} = \mathrm{cS}2/\mathrm{E}$) and light yields ($\mathrm{LY} = \mathrm{cS}1/\mathrm{E}$) allows for the reconstruction of a linear energy response of both S1 and S2 signals by rewriting Eq.~\ref{eq:energy_transformation} to:

\begin{equation}
\mathrm{CY} = -\frac{g_2}{g_1} \cdot \mathrm{LY} + \frac{g_2}{W}.
\label{eq:LY_CY_parametrization}
\end{equation}

\begin{figure*}
  \centering
  \includegraphics[width=\textwidth]{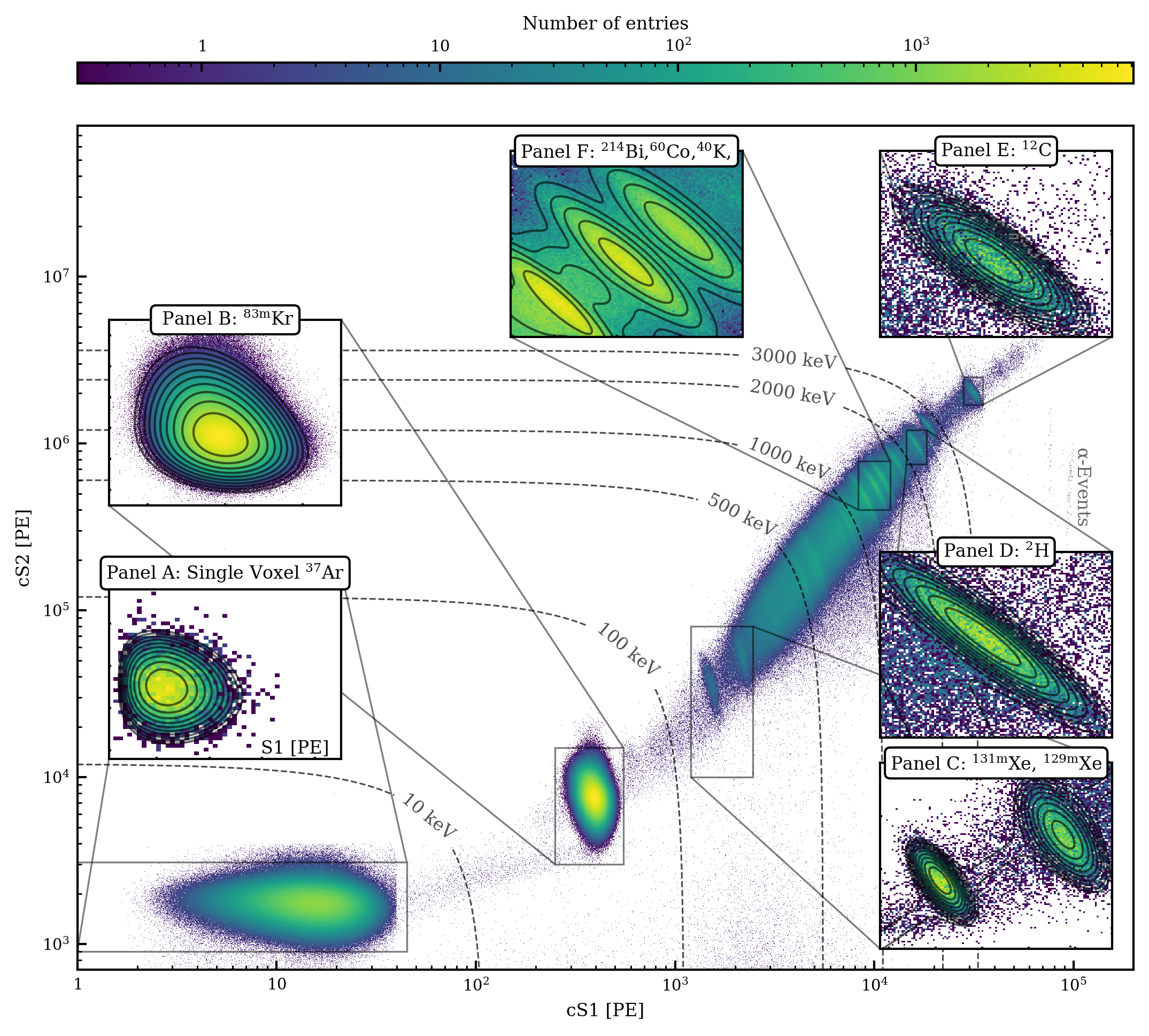}
  \caption{Two-dimensional histogram in cS1 and cS2 space covering most of the XENONnT energy range. The monoenergetic lines used in to extrct the $g_1$ and $g_2$ parameters are highlighted in subplots. The displayed data is a combination of multiple calibration datasets (\isotope[37]{Ar}, \isotope[83\text{m}]{Kr}, and \isotope[241]{AmBe}) as well as background data. The \isotope[37]{Ar} subplot displays the fit in a single voxel using the S1 and cS2 space. Events of the activated xenon lines \isotope[131\text{m}]{Xe} and \isotope[129\text{m}]{Xe} as well as \isotope[12]{C} and \isotope[2]{H} events are present in \isotope[241]{AmBe} calibration data. \isotope[60]{Co}, \isotope[214]{Bi} and \isotope[40]{K} can be found in background, \isotope[83\text{m}]{Kr} and \isotope[241]{AmBe} calibration data but only background and \isotope[83\text{m}]{Kr} calibration data is used for the corresponding fit.}
  \label{fig:energy_res:LCY_measurment}
\end{figure*}

The anti-correlation of light and charge signals outlined in Eq.~\ref{eq:energy_transformation} leads to mono-energetic lines appearing as rotated ellipses when plotting cS1 against cS2, as depicted in Fig.~\ref{fig:energy_res:LCY_measurment} for the full energy range of the merged calibration and blinded data. Rotated two-dimensional Gaussian functions are used to fit each monoenergetic ellipse and extract its corresponding $\mathrm{LY}$  and $\mathrm{CY}$. This method has been adapted to each mono-energetic source as described in the following:

\begin{itemize}
    \item For the K-shell \isotope[37]{Ar} peak at 2.82\;keV, being close to the S1 detection efficiency introduced in Sec.~\ref{subsec:rec_eff}, it is necessary to perform the fit in the uncorrected S1 and cS2 parameter space. The detection volume is segmented in different voxels, and the $L_c(\text{R}, \phi, \text{Z})$ correction shown in Eq.~\ref{eq:relative_S1_corr_det}, evaluated in each voxel barycenter, was manually applied to the S1 mean obtained. The S1 signal is modeled with a skew-Gaussian distribution, which has proven to be a more suitable model for $\mathcal{O}(1)$\;keV ERs~\cite{XENON1T:Ar37-calibration}, convolved with the data-driven estimation of the S1 detection efficiency. For cS2, a normal distribution is considered. An example of 2D fit in a single voxel with the projections is shown in Fig.~\ref{fig:energy_res:LCY_measurment} panel A. The average $\mathrm{LY}$ and $\mathrm{CY}$ over all the voxels, $\langle \mathrm{LY} \rangle = (5.3325 \pm 0.0014)$\;PE/keV and $\langle \mathrm{CY} \rangle = (320.4 \pm 0.3)$\;PE/keV, are used to derive the $g_1$ and $g_2$ parameters.
    \item The \isotope[83\text{m}]{Kr} peak, shown in Fig.~\ref{fig:energy_res:LCY_measurment} panel B, exhibits a tail toward larger cS2, most likely induced by non-perfect signal correction of field inhomogeneities. This artifact is also present in the other mono-energetic lines but has a negligible impact on the LY/CY measurement. A skew-Gaussian in cS2 is used to model the observed tail.
    \item Unlike the former two mono-energetic lines that come from dedicated calibrations, the \isotope[129\text{m}]{Xe} and \isotope[131\text{m}]{Xe} lines are present in the background data after a neutron calibration. As depicted in Fig.~\ref{fig:energy_res:LCY_measurment} panel C, they stand on top of a continuous background band from Compton scatter and $\upbeta$-decay. Therefore, a two-dimensional function featuring a linear profile along the cS1 axis, a Gaussian profile along the cS2 axis, and a rotation angle in the plane is incorporated into the rotated 2D Gaussian to improve the fit.
    \item Several high-energy gamma lines (above \SI{300}{\keV}), either originating from the radiogenic background of the detector materials or induced during \isotope[241]{AmBe} calibration, serve as additional reference lines. During \isotope[241]{AmBe} calibration, alpha capture on \isotope[9]{Be} creates a compound nucleus, \isotope[13]{C^*}, which rapidly decays by emitting a neutron. This process can lead to an excited state of \isotope[12]{C}, emitting a \SI{4.4}{\MeV} gamma. The neutron can also be captured by hydrogen in the water tank surrounding the TPC, leading to the emission of a \SI{2.2}{\MeV} gammas. Such high-energy gamma can reach the sensitive volume of the TPC as shown in Fig.~\ref{fig:energy_res:LCY_measurment} panels D and E.
    Additionally, radiogenic gamma lines from \isotope[60]{Co} with energies of \SI{1173.2}{\keV} and \SI{1332.5}{\keV}, and from \isotope[40]{K} at \SI{1460.8}{\keV}, originate from the detector materials and can be seen in background data (Fig.~\ref{fig:energy_res:LCY_measurment}, panel F). These high-energy gamma lines are fitted with rotated 2D Gaussian functions, taking into account continuous background contribution from Compton scattering and beta decay. However, these lines are not included in the final \(g_1/g_2\) fit, as stated previously. 
\end{itemize}

\begin{figure}[h!]
  \centering
  \includegraphics[width=\linewidth]{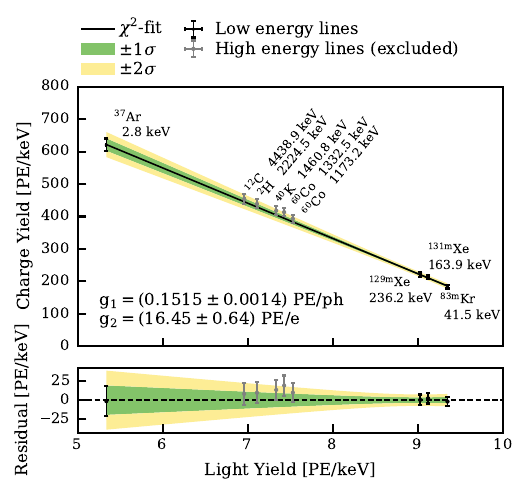}
  \caption{Anti-correlation between the measured light yield and charge yield using mono-energetic lines. The black solid line represents the best linear fit to the data obtained from \isotope[37]{Ar}, \isotope[83\text{m}]{Kr}, \isotope[129\text{m}]{Xe}, and \isotope[131\text{m}]{Xe} low-energy lines. Data points derived from high-energy lines, not included in the fit, are displayed as grey markers. Despite not being used for the fitting process, these high-energy points remain consistent with the fitted model within their uncertainties.}
  \label{fig:energy_res:doke_plot}
\end{figure}

Before computing the gain parameters $g_1$ and $g_2$, the measured LY and CY values are corrected for the energy-dependent peak reconstruction bias introduced in Sec.~\ref{sec3:peak_rec_bias}. This correction is applied only to the LY/CY measurement and not directly to the cS1/cS2 value, which results in a biased energy scale. To minimize this bias in the low-energy region, the peak reconstruction bias correction is rescaled to have zero bias for the \isotope[37]{Ar} line. The observed energy bias is characterized and incorporated into the inference as discussed later. A systematic error of 3.2\% on the CY is used to account for the average distance between the best-fit prediction using the four low-energy lines and the high-energy lines. Fig.~\ref{fig:energy_res:doke_plot} shows the relation between measured CY and LY. Factors $g_1$ and $g_2$ are extracted using a linear fit following Eq.~\ref{eq:LY_CY_parametrization}. The parameters extracted from the fit that allows us to build our ER energy scale are $g_1 = (0.151 \pm 0.001)\; \mathrm{PE} / \text {ph}$ and $g_2 = (16.5 \pm 0.6)\;\mathrm{PE} / \text{e}^{-}$.

\begin{figure*}
  \centering
  \includegraphics[width=\textwidth]{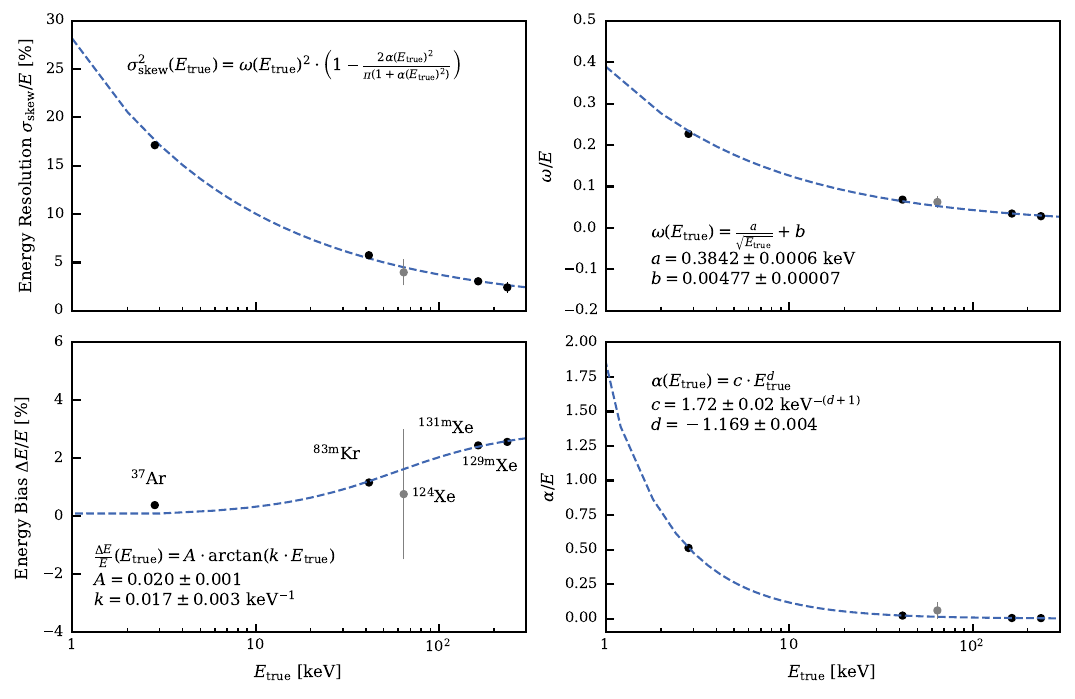}
  \caption{The top left panel shows the measured energy resolution ($\sigma/\mu$) from the low-energy lines used in the $g_1$/$g_2$ fit (c.f. Fig.\ref{fig:energy_res:doke_plot}). An empirical model, specifically tailored for the skew-Gaussian fitting method, characterizes the energy resolution, which depends on the width ($\upomega$) and skewness ($\upalpha$) parameter measurements. These parameters and their corresponding fits are displayed in the top right and bottom right panels, respectively, with the fitted parameter values annotated. The bottom left panel illustrates the relative energy bias, modeled with an empirical function, and its fitted parameters. The gray point (not included in the fits) shows the dominant KK-capture peak from \isotope[124]{Xe} $2\nu \mathrm{ECEC}$ decay, which was recently observed for the first time in the XENON1T experiment~\cite{PhysRevC.106.024328}. The energy resolution and the relative energy bias models are propagated to the main analyses by reshaping the expected energy spectra. }
  \label{fig:energy_res:energy_resolution}
\end{figure*}

The reconstructed energy of mono-energetic lines is fitted with a free skew-Gaussian function and a free linear background to model the detector's energy resolution. The skew-Gaussian model better describes the mis-modeling and imperfect signal correction of low-energy lines mentioned earlier. This model consists of three parameters: the width ($\upomega$), the location ($\xi$), and the skewness ($\upalpha$), which together allow the reconstruction of the skew mean ($\upmu_{\mathrm{skew}}$) and standard deviation ($\upsigma_{\mathrm{skew}}$). The top left panel of Fig.~\ref{fig:energy_res:energy_resolution} shows the measured energy resolution ($\sigma / \mu$) for the four low-energy lines used in the \(g_1/g_2\) fit, along with the empirical model fitted to these data points. The top right and bottom right panels display the width and skewness parameters, respectively, each with their corresponding fitted values annotated used to model the energy resolution. The skew-Gaussian model transitions to a regular Gaussian distribution for high-energy lines as the skewness converges to zero. The relative energy bias, shown in the bottom left panel of Fig.~\ref{fig:energy_res:energy_resolution}, is characterized using an empirical function in the reconstructed energy space and propagated into the analyses by reshaping the expected energy spectra. Additionally, the dominant KK-capture peak from \isotope[124]{Xe} $2\nu \mathrm{ECEC}$ decay at \SI{64.3}{\keV}~\cite{PhysRevC.106.024328} is shown as a cross-check of the energy reconstruction. The significant decrease in the background level makes this peak distinctly prominent, thereby rendering it particularly suitable for cross-checking the accuracy of the energy reconstruction process.

\section{Conclusion and outlook}
\label{sec:conclusion}
This paper presents the data analysis techniques employed for the WIMPs~\cite{xenon_nt_wimp.131.041003} and low-energy electronic recoil~\cite{lowER:Aprile:2020tmw} searches during the first XENONnT science run. It details the processes of signal reconstruction and correction, event building, selection criteria, and energy estimation. The majority of the methodologies outlined are applicable to ongoing and forthcoming searches for WIMP, alternative dark matter hypotheses, and various low-background investigations.

Throughout the entire science run, the detector functioned under consistent conditions. The TPC photosensors exhibited stability and reliability in their response throughout the commissioning phase and first science run. A mere 3\% of the PMTs were deemed non-operational, corresponding to a failure rate of approximately a factor of 5 lower compared to XENON1T~\cite{Aprile:2019bbb}. The light and charge yield responses were stable throughout the entire science data taking, with fluctuation smaller than 1\% and 1.9\%, respectively. The improvement in the xenon purification allows us to reach unprecedented low concentrations of electronegative contaminants, thanks to which our electron lifetime was constantly above ~10\;ms. This is a factor 10 improvement with respect to XENON1T~\cite{Aprile:2019bbb}.

Regular calibrations using an $^{83\text{m}}$Kr internal source were employed to assess the TPC's response and to calculate signal corrections. These corrections accounted for detector artifacts, such as distortions in the electric drift field and spatial inhomogeneity in detecting and reconstructing S1 and S2 signals. Additionally, internal sources of $^{220}$Rn and $^{37}$Ar were utilized to characterize the electronic recoil response. An external source of $^{241}$AmBe was employed to assess the TPC response to nuclear recoil events and to evaluate the detection efficiency of the neutron veto~\cite{xenonnt_nv}. For additional information regarding the characterization of the ER and NR detector responses, we recommend that readers consult~\cite{xenonnt_analysis_2}. 

A novel data processing software has been developed for the new XENONnT triggerless data taking~\cite{XENON:2022vye}. Its performance are optimized based on full-chain waveform simulation, thanks to which the peak finding efficiencies, peak reconstruction, and event reconstruction biases are also estimated. Improved MC optical simulation of the TPC is employed to tune position reconstruction algorithms, and detailed electric field simulation is used to improve the homogeneity of the drift field and the understanding of the ionization signal~\cite{XENON:xenonnt_field_cage}. The selection criteria resemble those used in XENON1T~\cite{Aprile:2019bbb}, with the addition of new criteria based on machine learning techniques that have further enhanced the data quality. The more advanced analysis with respect to its predecessor, together with the hardware improvements for the background reduction (e.g., Radon removal system), has made possible the measurement of the lowest background below 30\,keV for a dark matter detector, equivalent to (15.8 $\pm$ 1.3)\;events/(tonne$\cdot$year$\cdot$keV) consisting of a factor 5 reduction concerning XENON1T~\cite{lowER:Aprile:2020tmw}.

The anti-correlation between charge and light yields has been confirmed for energy depositions ranging from a few keV to the MeV scale. The calibration of the energy scale was achieved by utilizing the $g_1 = (0.151 \pm 0.001)\; \mathrm{PE} / \text {ph}$ and $g_2 = (16.5 \pm 0.6)\;\mathrm{PE} / \text{e}^{-}$ detector parameters and by integrating both direct and proportional scintillation signals. The resulting energy resolution was found to be on par with XENON1T~\cite{Aprile:2020yad}.

Although the foundational elements of the XENONnT analysis pipeline are well-established, ongoing efforts are dedicated to enhancing it to improve detector performance further and deepen our understanding of its response. 
We are working towards understanding the origin of accidental coincidence backgrounds and exploring methods for their further reduction. Additionally, new calibration sources (e.g., $^{232}$Th and YBe~\cite{xenonnt_ybe_2024}) have been investigated to enhance the characterization of the detector's ER and NR responses. On the hardware front, ongoing improvements to the subsystem aim to increase xenon purity, which will benefit subsequent analyses. Lastly, the simulation framework is being refactored to adopt a more modular approach and to improve the physical description of the detector, such as electron diffusion below the perpendicular wires.
XENONnT is actively collecting new data, and advancements in hardware have significantly boosted its performance. The physics reach of XENONnT is broad, encompassing a variety of novel and compelling analyses currently in progress.

\begin{acknowledgments}

We gratefully acknowledge support from the National Science Foundation, Swiss National Science Foundation, German Ministry for Education and Research, Max Planck Gesellschaft, Deutsche Forschungsgemeinschaft, Helmholtz Association, Dutch Research Council (NWO), Fundacao para a Ciencia e Tecnologia, Weizmann Institute of Science, Binational Science Foundation, Région des Pays de la Loire, Knut and Alice Wallenberg Foundation, Kavli Foundation, JSPS Kakenhi and JST FOREST Program ERAN in Japan, Tsinghua University Initiative Scientific Research Program, DIM-ACAV+ Région Ile-de-France, and Istituto Nazionale di Fisica Nucleare. This project has received funding/support from the European Union’s Horizon 2020 research and innovation program under the Marie Skłodowska-Curie grant agreement No 860881-HIDDeN.
We gratefully acknowledge support for providing computing and data-processing resources of the Open Science Pool and the European Grid Initiative, at the following computing centers: the CNRS/IN2P3 (Lyon - France), the Dutch national e-infrastructure with the support of SURF Cooperative, the Nikhef Data-Processing Facility (Amsterdam - Netherlands), the INFN-CNAF (Bologna - Italy), the San Diego Supercomputer Center (San Diego - USA) and the Enrico Fermi Institute (Chicago - USA). We acknowledge the support of the Research Computing Center (RCC) at The University of Chicago for providing computing resources for data analysis.
We thank the INFN Laboratori Nazionali del Gran Sasso for hosting and supporting the XENON project. 

\end{acknowledgments}

\bibliographystyle{spphys}
\bibliography{main}

\end{document}